\newif\ifpreprint
\preprinttrue
\ifpreprint 
\documentclass[twocolumn,a4paper,natbib]{article}
\usepackage[affil-it]{authblk}
\usepackage{amssymb,amsmath}
\usepackage{graphicx}
\usepackage[cm]{fullpage}
\else 
\documentclass[times,a4paper]{qjrms4}
\fi 

\usepackage{color}
\usepackage{stmaryrd} 

\usepackage[colorlinks,bookmarksopen,bookmarksnumbered,citecolor=red,urlcolor=red]{hyperref}
\usepackage{algorithmic}
\usepackage{algorithm}

\newcommand{\laplhoriz}{\Delta_{2d}}
\newcommand{\nablahoriz}{\nabla_{2d}}

\newcommand{\DDt}[1]{\frac{D#1}{Dt}}

\newcommand{\ddr}[1]{\frac{\partial #1}{\partial r}}
\renewcommand{\vec}[1]{{\boldsymbol{#1}}}
\newcommand{\dt}{\Delta t}
\newcommand{\adt}{\alpha\dt}
\newcommand{\omadt}{(1-\alpha)\dt}

\newcommand{\order}[1]{\mathcal{O}(#1)}

\newcommand{\Us}{\operatorname{s}}

\newcommand{\Rearth}{R_{\operatorname{earth}}}
\newcommand{\bg}{{*0}}

\newcommand{\titer}{\ensuremath{t_{\operatorname{iter}}}}
\newcommand{\tsolve}{\ensuremath{t_{\operatorname{solve}}}}
\newcommand{\tsetup}{\ensuremath{t_{\operatorname{setup}}}}
\newcommand{\ttotal}{\ensuremath{t_{\operatorname{total}}}}

\begin{document}

\title{Massively parallel solvers for elliptic PDEs in Numerical Weather- and Climate Prediction}
\ifpreprint 
\author[*,1]{Eike M\"uller}
\author[1]{Robert Scheichl}
\affil[1]{Department of Mathematical Sciences, University of Bath, Bath BA2 7AY, United Kingdom}
\affil[*]{Email: \texttt{e.mueller@bath.ac.uk}}
\renewcommand\Authands{ and }
\else 
\runningheads{E. H. M\"{uller}, R. Scheichl}{Scalability of Elliptic Solvers in NWP}
\author{Eike H. M\"{u}ller\affil{a}\corrauth\ and Robert Scheichl\affil{a}}
\address{\affilnum{a}Department of Mathematical Sciences, University of Bath, Bath BA2 7AY, United Kingdom}
\corraddr{\texttt{e.mueller@bath.ac.uk}}
\fi 
\ifpreprint 
\twocolumn[
\begin{@twocolumnfalse}
\maketitle
\fi 
\begin{abstract}
The demand for substantial increases in the spatial resolution of global weather- and climate- prediction models makes it necessary to use numerically efficient and highly scalable algorithms to solve the equations of large scale atmospheric fluid dynamics. {For stability and efficiency reasons several of the operational forecasting centres, in particular the Met Office and the ECMWF in the UK, use semi-implicit semi-Lagrangian time stepping in the dynamical core of the model. The additional burden with this approach is that a three dimensional elliptic partial differential equation (PDE) for the pressure correction has to be solved at every model time step and this often constitutes a significant proportion of the time spent in the dynamical core.}
To run {within} tight operational time scales the solver has to be parallelised and there {seems to be} a (perceived) misconception that elliptic solvers do not scale to large processor counts and hence implicit time stepping can not be used in very high resolution global models.
After reviewing several methods for solving the elliptic PDE for the pressure correction and their application in atmospheric models we demonstrate the performance and very good scalability of Krylov subspace solvers and multigrid algorithms for a representative model equation with more than $\boldsymbol{10^{10}}$ unknowns on 65536 cores on HECToR, the UK's national supercomputer. 
For this we tested and optimised solvers from two existing numerical libraries ({\tt DUNE} and {\tt hypre}) and implemented both a Conjugate Gradient solver and a geometric multigrid algorithm based on a tensor-product approach which exploits the strong vertical anisotropy of the discretised equation. We study both weak and strong scalability and compare the absolute solution times for all methods; in contrast to one-level methods the multigrid solver is robust with respect to parameter variations.
\end{abstract}
\ifpreprint 
\else 
\maketitle
\fi 
\ifpreprint 
\end{@twocolumnfalse}
\vspace{3ex}
]
\fi 

\section{Introduction}
Modern forecast models in numerical weather- and climate- prediction (NWP) use the fully compressible non-hydrostatic Navier-Stokes equations to simulate the dynamics of the atmosphere. If the atmospheric fields are advanced forward in time by explicit time stepping, there are severe limitations on the size of the model timestep, which --- for the model to remain stable---, must not exceed the ratio of the grid size and velocity of the fastest waves. For a compressible fluid these are acoustic waves with a speed of several hundred metres per second at ground level.
Even if the vertical propagation of sound waves can be dealt with, horizontal grid resolutions {are likely to} be reduced to a few kilometres in the future, severely limiting the model time step. In contrast, implicit time stepping allows to run the model with larger time steps without compromising its stability and without distorting the large scale flow close to geostrophic balance.
However, this requires the solution of an elliptic partial differential equation (PDE) for the pressure correction at every time step. In a non-hydrostatic model this equation has to be solved in three dimensions, i.e. on a spherical shell representing the earth's atmosphere.\vspace{2ex}

\noindent {\em Elliptic PDEs in semi-implicit time stepping.}
Schematically, for a set of
 variables $\phi=(u,v,w,\pi,\theta,\dots)$ the time evolution is described by the equation 
\begin{eqnarray}
  \frac{D\phi(\vec{x},t)}{Dt} &=& \mathcal{N}[\phi(\vec{x},t)] + R_\phi.\label{eqn:TimeEvolution}
\end{eqnarray}
Here $D/Dt$ is the material derivative; the (not necessarily linear) operator $\mathcal{N}$ describes the large scale physical processes such as the Coriolis force, pressure gradients, gravitational acceleration and divergence due to mass fluxes. In the dynamical core of the  model, unresolved sub-gridscale processes, such as turbulence, convection and thermodynamic phase transitions, are included as external forcings represented by the term $R_\phi$.

In the semi-Lagrangian formulation, first introduced by \cite{Robert1981}, the advection terms $D\phi/Dt$ are evaluated as {differences of fields} at the next time step and {at the departure point of the current time step}. The other terms are treated semi-implicitly following \cite{Kwizak1971}.
The semi-Lagrangian semi-implicit time discretisation scheme 
was first applied to a fully non-hydrostatic model in \cite{Tanguay1990}; a review of more recent models which use semi-implicit semi-Lagrangian time stepping can be found in \cite{Steppeler2003}. Calculation of the fields $\phi_{\mathbf{j}+1}
$ at the next timestep {$t_{\mathbf{j}+1}$} requires the solution of an elliptic PDE for the pressure correction $\pi'_{\mathbf{j}+1}$.
After discretisation and linearisation, this PDE can be written as a large algebraic problem
\begin{eqnarray}
  A\pi'_{\mathbf{j}+1}
&=&f_{\mathbf{j}}
  \label{eqn:AlgebraicProblem}
\end{eqnarray}
where the pressure correction at the next time step {$t_{\mathbf{j}+1}$} is represented by an $n$ dimensional solution vector $\pi'_{\mathbf{j}+1}
$ and the right hand side $f_{\mathbf{j}}
$ only depends on fields at the current time step {$t_{\mathbf{j}}$}. The matrix $A$ is a sparse $n\times n$ matrix from the discretisation of the continuum operator. 

The number of degrees of freedom $n$ is very large. To see this, note that the number of grid cells of area $h^2$ necessary to cover the surface of the earth is $n_{2d}=4\pi \Rearth^2/h^2$, which gives $n_{2d}\approx 5\cdot10^{8}$ for a grid spacing of $h=1km$. The typical number of cells in the vertical direction is of the order of $n_z\approx 100$ resulting in a total number of degrees of freedom of $n \ge 10^{10}$.

On the other hand, the following estimate reveals the high demands on the performance of the solver of the elliptic problem in (\ref{eqn:AlgebraicProblem}):
for a horizontal resolution of $1km$, the limitations on the explicit time step are \mbox{$\Delta t\lesssim \Delta x/c_s\approx 3s$} where $c_s\approx 350ms^{-1}$ is the speed of sound at ground level. By using implicit time stepping, this can be extended tenfold to around $30s$. To produce a 5 day global forecast this requires $14400$ {time steps}.
When the model is run operationally, the time available for the {dynamical core is typically less than an hour, often allowing less than twenty minutes for the elliptic solver.} In total, this means that the non-linear equation has to be solved in less than {$0.1s$}. Usually this requires a very small number of iterations (around 3) of the Newton algorithm, in each of which the linear PDE has to be solved. Hence, the time available for one linear solve is around $0.03s$ (requiring terascale computing capability).\vspace{2ex}
\noindent {\em Efficient and massively scalable solvers.}
To solve a problem of this size in operational time frames requires state-of-the-art iterative solvers, such as suitably preconditioned Krylov subspace or multigrid methods. The algorithms have to scale algorithmically to problem sizes of $\order{10^{10}}$, i.e. the number of iterations should not grow significantly with an increase in resolution. They should also be stable with respect to variations of the coefficients. For optimal performance it is crucial to exploit the strong vertical coupling in the discretised operator.

Problems of this size can only be solved in a reasonable time on massively parallel computers.
{This} introduces additional complications such as communication overheads, synchronisation- and load balancing issues.
In this paper we intend to dispel the common misconception that solvers for elliptic PDEs arising from semi-Lagrangian semi-implicit time stepping do not scale {to large core counts}.

To demonstrate this we compare different solvers and study their performance and scalability for the solution of a model equation on up to 65536 cores on HECToR, the UK's national supercomputer which is hosted and managed by the Edinburgh Parallel Computing Centre (EPCC). We tested and optimised existing solvers from the Iterative Solver Template Library (ISTL) which is part of the Distributed and Unified Numerics Environment (DUNE; \cite{Bastian2008a,Bastian2008b}) and {from} the hypre library (\cite{FalgoutYang2002,Falgout2006}). The most efficient general purpose preconditioners from these packages are algebraic multigrid solvers, {which} have been shown to scale to 100,000s of cores  for different problems before (\cite{IppischBlatt2011,Hypre2012a,Hypre2012b}).
However, algebraic multigrid methods have an additional cost for setting up the discretisation matrix and constructing the hierarchy of multigrid levels. Since the problem we are studying is discretised on a grid which can be written as the tensor-product of a (semi-)structured horizontal mesh and a regular one dimensional grid in the vertical direction, we also implemented a geometric multigrid code based on the tensor-product idea in \cite{BoermHiptmair1999}.

All the solvers show very good weak scaling to up to 65536 cores and they are {all} algorithmically {scalable}.
The multigrid solvers require substantially less iterations than the Conjugate Gradient (CG) algorithm, preconditioned with vertical line relaxation, {and they are fully robust under variations of the parameters in the model equation (as opposed to CG)}. In terms of absolute performance, {not surprisingly,} the matrix-free geometric multigrid solver outperforms the algebraic multigrid solvers as (i) it requires less iterations to converge, (ii) each iteration is around twice as fast and (iii) it does not have any coarse-level setup costs. {Our numerical results show that the difference is quite significant (more than factor 10).} We are able to carry out one multigrid V-cycle for a problem with $3.4\cdot 10^{10}$ degrees of freedom in $0.177s$ on 65536 cores, giving a total solution time of around one second to reduce the residual by five orders of magnitude. {We also demonstrated good strong scaling for different problem sizes. The
tests show that it is realistic that the total solution time can be decreased below the threshold required for operational runs. {In addition to a bespoke Fortran implementation for regular horizontal grids, the geometric multigrid code has also been implemented in the DUNE framework, which allows the treatment of more general (semi-) structured horizontal grids and future extensions to alternative discretisation schemes (such as higher-order discontinuous Galerkin, cf.~\cite{Bastian2012}).}\vspace{2ex}

\noindent {\em Structure.}
This paper is organised as follows: the idea of semi-implicit semi-Lagrangian time stepping is introduced in Section~\ref{sec:SISLTimeStepping} and the most important features of the resulting elliptic model equation are discussed in Section~\ref{sec:HelmholtzEquation}. Modern methods for solving elliptic PDEs and their applications in numerical weather- and climate-prediction are reviewed in Section~\ref{sec:SolverReview}. The solver libraries which were used for this work, and the implementation of the matrix-free Krylov subspace solver and geometric multigrid algorithm tailored towards the model problem are described in Section \ref{sec:Implementation}. Results for the absolute performance as well as weak- and strong- scaling tests are reported in Section \ref{sec:Results}, where we also study the robustness of the multigrid solver. Our conclusions are summarised in Section \ref{sec:Conclusions}.
\ifpreprint
More technical details have been relegated to the appendices; the full derivation of the model equation from the fundamental equations of atmospheric fluid dynamics can be found in Appendix \ref{sec:DerivationModelEquation}. For reference some of the most important numerical algorithms discussed in this article are collected in Appendix \ref{sec:Algorithms}.
\fi 
\section{Semi-implicit semi-Lagrangian time stepping}
\label{sec:SISLTimeStepping}
As outlined in the introduction, in the semi-implicit semi-Lagrangian (SISL) time stepping scheme the advection terms in the Navier-Stokes equations are handled by calculating the difference {between the field at point $\vec{x}$ and} at time {$t_{\mathbf{j}+1} = t_{\mathbf{j}}+\Delta t$} and the field at the previous time step {$t_{\mathbf{j}}$}, evaluated at the departure point {$\vec{x}_D$} (which can be calculated from the velocity field). For a generic material 
derivative\vspace{-2ex}
\begin{eqnarray} 
  \DDt{\phi(\vec{x},t)} &=& \mathcal{N}[\phi(\vec{x},t)] + R_\phi
\end{eqnarray}
this amounts to
\begin{eqnarray}
\left[\phi-\adt\;\mathcal{N}[\phi]\right]_{\mathbf{j}+1}
&& \hspace*{-0.7cm}(\vec{x})  \label{eqn:SISLScheme}\\ 
=&&\hspace*{-0.5cm}\left[\phi+\omadt\; \mathcal{N}[\phi]+R_\phi\right]_{\mathbf{j}}
(\vec{x}_D)\notag
\end{eqnarray}
where the off-centering parameter $\alpha$ describes the ``implicitness'' with which the terms in $\mathcal{N}[\phi]$ are treated in the scheme ($\alpha=1$ corresponds to {implicit Euler} and $\alpha=0$ to {explicit Euler}; for $\alpha=\tfrac{1}{2}$ it reduces to the scheme described in \cite{CrankNicolson1996}).

To illustrate the method, consider the (2D) shallow water equations\footnote{The dimensionless fields $\vec{v}$ and $\eta$ are obtained from the physical fields by rescaling with $c_g$ and the (constant) depth $\Phi$.} for the velocity $\vec{v}$ and height perturbation $\eta$
\begin{eqnarray}
  \frac{D\vec{v}(\vec{x},t)}{Dt} &=& -c_g \nabla\eta(\vec{x},t),\\
  \frac{D\eta(\vec{x},t)}{Dt} &=& -c_g(1+\eta(\vec{x},t)) \nabla\cdot \vec{v}(\vec{x},t).
\end{eqnarray}
The gravity wave velocity is $c_g = \sqrt{g\Phi}$. Following the semi-implicit semi-Lagrangian time stepping scheme (\ref{eqn:SISLScheme}), these equations can be {semi-}discretised in time as 
\begin{eqnarray}
  \left[\vec{v} + \alpha \dt\;c_g \nabla\eta\right]_{\mathbf{j}+1}(\vec{x}) &=& {\vec{f}^{\vec{v}}_{\mathbf{j}}}(\vec{x}_D), \label{eqn:SISLSW_1}\\
\left[\eta + \alpha \dt\;c_g (1+\eta)\nabla\cdot \vec{v}\right]_{\mathbf{j}+1}(\vec{x}) &=& {f^{\eta}_{\mathbf{j}}}(\vec{x}_D)
  \label{eqn:SISLSW_2}
\end{eqnarray}
where all terms evaluated at the time step {$t_{\mathbf{j}}$ are collected in $\vec{f}_{\mathbf{j}}^{\vec{v}}(\vec{x}_D)$ and $f_{\mathbf{j}}^{\eta}(\vec{x}_D)$} on the right hand side.
By taking the divergence of (\ref{eqn:SISLSW_1}) and inserting it into (\ref{eqn:SISLSW_2}) it is easy to see that (after linearisation) one arrives at the following elliptic PDE for the height variation $\eta$ at the next timestep
\begin{eqnarray}
  \left[
    -\omega_\mathrm{SW}^2 \laplhoriz \eta + \eta
  \right]_{\mathbf{j}+1}(\vec{x}) &=& {f_{\mathbf{j}}^\mathrm{SW}}(\vec{x}_D)
\end{eqnarray}
where the right hand side ${f_{\mathbf{j}}^\mathrm{SW}}$ only depends on fields at the current time step. Here $\laplhoriz$ is the Laplace operator in two dimensions. Note that the relative size of the second order term $\laplhoriz \eta$ is given by the parameter
\begin{eqnarray}
  \omega_\mathrm{SW}^2 &=& \left(\alpha c_g \Delta t\right)^2
\end{eqnarray}
{which decreases} quadratically with the time step $\Delta t$.

The full equations of a three dimensional non-hydrostatic model can be derived analogously (see \cite{Wood2013}).
The resulting elliptic problem is the following three dimensional PDE for the Exner pressure correction $\pi'$:
\begin{eqnarray}
  {\left[-\omega_\mathrm{3D}^2 \left(\laplhoriz
  + D^{(z)}
   \right)\pi'+\gamma \pi'\right]_{\mathbf{j}+1}(\vec{x}) \ = \ f_{\mathbf{j}}(\vec{x}_D).} \label{eqn:FullSISLEquation}
\end{eqnarray}
{where $\omega_\mathrm{3D}^2 = (\adt)^2 c_p\theta^\bg\pi^\bg$ and the second order differential operator in the radial direction is}
\begin{eqnarray}
D^{(z)}X = {\frac{1}{\left(\pi^\bg\right)^\gamma c_p\theta^\bg r^2}}\,\frac{\partial}{\partial r}\left(\frac{\left(\pi^\bg\right)^\gamma c_p\theta^\bg {r^2}}{1+(\adt)^2\left(N^\bg\right)^2}\frac{\partial X}{\partial r}\right).\notag
\end{eqnarray}
Here, $\laplhoriz$ denotes the Laplacian in the horizontal direction on the surface of the sphere and $\gamma = \frac{c_p-R_d}{R_d}$, where $c_p$ and $R_d$ are the specific heat capacity and the specific gas constant of dry air. The fields $\theta^\bg$ and $\pi^\bg$ are background profiles for the potential temperature and Exner pressure which only depend on the vertical {(radial)} coordinate. 

\section{Elliptic model equation}\label{sec:HelmholtzEquation}
For the scaling tests in this article we do not include the vertical profiles but set them to a constant value of 1, {since (after rescaling of the problem) they only appear in the zero-order term and in the vertical operator $D^{(z)}$ and we believe that by construction all our solvers are robust to this generalisation. Further tests are required to confirm this.} We also rescale all {dimensional} quantities such that the radius of the earth is 1. This leads to the following positive definite elliptic model problem
  \begin{equation}
    -\omega^2 \left(\laplhoriz u+\lambda^2  \frac{1}{r^2}\frac{\partial}{\partial r}\left(r^2 \frac{\partial u}{\partial r}\right) \right)+u \;=\; f
  \label{eqn:ModelEquation}
  \end{equation}
which is solved for $u$ in the spherical shell defined by $1\le r \le1+H$. Here $H=D/\Rearth=\tfrac{1}{100}$ is the ratio between the depth of the atmosphere and the radius of the earth.
The parameters $\omega^2$ and $\lambda^2$ are
\begin{xalignat}{2}
  \omega^2 &= \left(\frac{\alpha c_h\Delta t}{\Rearth}\right)^2, &
  \lambda^2 &= \frac{1}{1+(\adt)^2\left(N^\bg\right)^2},
\label{eqn:omega2lambda2}
\end{xalignat}
where $c_h=550ms^{-1}$ is the velocity of the fastest waves in the system.
\ifpreprint
As described in appendix \ref{sec:DerivationModelEquation}, this
\else 
This
\fi 
velocity is related to the speed of horizontal acoustic waves at ground level, $c_s\approx 350ms^{-1}$, by a factor or order one, i.e. $c_h^2=\gamma c_s^2$ with $\gamma=\frac{c_p-R_d}{R_d}=2.506$.
The buoyancy frequency in equation (\ref{eqn:omega2lambda2}) is given by $N^\bg=0.018s^{-1}$. $\Rearth=6371km$ is the radius of the earth and we choose $\alpha=0.5$ for the off-centering parameter. 
Homogeneous Neumann boundary conditions $\tfrac{\partial u}{\partial r}=0$ are used at the bottom and top of the atmosphere.

Note that in contrast to the Poisson equation, the solution of (\ref{eqn:ModelEquation}) is unique even if homogeneous Neumann boundary conditions are used on all external surfaces. In particular, if $\vec{\xi}$ denotes tangential coordinates on the surface of the sphere, the solution can be written as
\begin{eqnarray}
  u(\vec{\xi},r) &=& \overline{u}(\vec{\xi}) + \delta u(\vec{\xi},r)
\end{eqnarray}
where $\overline{u}(\vec{\xi})$ does not depend on the radial coordinate $r$, and is referred to as a \textit{vertical zero mode} as it is annihilated by the vertical derivative, $\tfrac{1}{r^2}\tfrac{\partial}{\partial r}\left(r^2\tfrac{\partial\overline{u}}{\partial r}\right)=0$. This mode is absent if homogeneous Dirichlet boundary condition are used, and, depending on the values of $\omega^2$ and $\lambda^2$, this can make the problem significantly better conditioned. It is for this reason that we use homogeneous Neumann 
boundary conditions in this work.
\subsection{Choice of grid and discretisation}
A plethora of grids can be used to discretise the surface of a sphere, and choosing the optimal grid for the dynamical core of a global forecast model is a problem in itself, see \cite{Staniforth2012} for a review. As discussed in detail in \cite{Buckeridge2010} one of the problems of a simple latitude-longitude grid is the convergence of grid lines at the poles and the resulting horizontal anisotropy which has a negative impact on the performance of the solver. This grid has other problems for parallelisation: near the pole {large communication stencils} and large halos are necessary to account for the transport of fields.
\begin{figure}
  \begin{center}
  \includegraphics[width=0.6\linewidth]{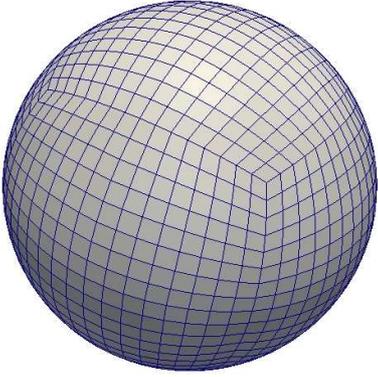}
  \caption{Cubed sphere grid}
  \label{fig:GridCubedSphere}
  \end{center}
\end{figure}
To avoid these problems and to still use a relatively simple grid we implemented the cubed sphere grid with a gnomonic mapping first discussed in \cite{Sadourny1972} (see Fig. \ref{fig:GridCubedSphere}). For each of the six faces a point in the spherical shell is constructed as follows:
\begin{eqnarray}
  \begin{pmatrix}
    x(r,\xi_1,\xi_2) \\ y(r,\xi_1,\xi_2) \\ z(r,\xi_1,\xi_2)
  \end{pmatrix}
  =
  \begin{pmatrix}
    r\sin(\theta(\xi_1,\xi_2)) \\ r\cos(\theta(\xi_1,\xi_2))\sin(\phi(\xi_2)) \\ r\cos(\theta(\xi_1,\xi_2))\cos(\phi(\xi_2))
  \end{pmatrix}\label{eqn:x1x2_1}
\end{eqnarray}
with
\begin{xalignat}{2}
  \tan(\phi(\xi_2)) &= \xi_2, &
  \tan(\theta(\xi_1,\xi_2)) &= \xi_1/{\textstyle \sqrt{1+\xi_2^2}}\label{eqn:x1x2_2}
\end{xalignat}
where $\xi_1,\xi_2\in[-1,1]$ and $r\in[1,1+H]$. A uniform grid with $n_x$ cells in each direction is then used for $\xi_1,\xi_2$, i.e. the horizontal grid spacing in these coordinates is $\Delta\xi=2/n_x$. For this projection the grid is non-orthogonal. Note, however, that in contrast to the conformal projection in \cite{Rancic1996}, the ratio of the size of the largest and smallest grid cell is bounded and no horizontal anisotropy or pole singularities arise in the limit {as $n_x \to \infty$}.

For simplicity the scaling runs reported in this article are carried out on one of the faces of the cubed sphere grid. Some runs on the entire sphere for both a cubed sphere grid and an icosahedral grid are reported in Section \ref{sec:ResultsDUNEGrid}.

In the vertical direction the grid is defined by a set of levels $r_k$ such that $1=r_0<r_1,\dots<r_{n_z}=1+H$. The grid spacing increases linearly with height, i.e. we set
\begin{eqnarray}
  r_k &=& 1 + H \left(\frac{k}{n_z}\right)^2\qquad\text{for $k=0,\dots,n_z$}.
  \label{eqn:VerticalGrading}
\end{eqnarray}
Having a smaller grid spacing near the earth surface is desirable in numerical weather and climate prediction to better resolve the flow in the lower layers of the atmosphere.

For the work in this paper we use a simple cell-centred finite volume discretisation. This amounts to approximating the fluxes through the surfaces of each cell in the grid by finite differences. The obtained stencil only involves the nearest neighbours and has a size of 7 for a rectangular grid. {The qualitative results of our scaling tests should not depend on this special structure, but this would require further tests.}

{The following two properties of the model equation (\ref{eqn:ModelEquation})} are crucial for the construction of an efficient solver.

\subsubsection{Vertical anisotropy:}
\label{sec:VerticalAnisotropy}
As the radius of the earth is much larger than the thickness of the atmosphere, after discretisation the operator in (\ref{eqn:ModelEquation}) contains a very strong anisotropy in the vertical direction. The relative size of the vertical derivative relative to the horizontal Laplacian can be estimated by\vspace{-1.5ex}
\begin{eqnarray}
  {\beta} &\approx& \lambda^2 \Big(\frac{\Delta x}{\Delta z}\Big)^2.
\end{eqnarray}
For not too small horizontal grid spacings the anisotropy {$\beta$} is significantly larger than one, i.e. the problem is highly anisotropic, and so vertical line relaxation (see Section \ref{sec:Preconditioning}) is highly efficient, either as a preconditioner in Krylov subspace methods or as a smoother in multigrid iterations, {as} demonstrated for example in \cite{Skamarock1997,Thomas1997,BoermHiptmair1999,Buckeridge2010}.

Note that due to the grading in (\ref{eqn:VerticalGrading}) the vertical grid spacing varies with height, so the relative strength of the horizontal and vertical couplings can be different at the bottom and the top of the atmosphere for very small $\Delta x$. It is, however, always grid aligned, so the theory in \cite{BoermHiptmair1999} can still be used to construct an efficient geometric multigrid solver.

\subsubsection{Horizontal coupling:}
\label{sec:HorizontalCoupling} 
In addition to the second order derivative terms the operator in (\ref{eqn:ModelEquation}) contains a zero order term. The importance of this term has already been pointed out in \cite{Leslie1973} and \cite{Hess1997}, who study the performance of multigrid solvers for the two dimensional Helmholtz equation arising from implicit time stepping in a hydrostatic model.

After discretisation the relative size of the horizontal derivative and the zero order term is controlled by the ratio of time step size and the spatial resolution, in particular it can be shown that the strength of the horizontal coupling (i.e. the size of the off-diagonal matrix entries\footnote{Off-diagonal matrix entries in the vertical direction can be ignored for this argument if vertical line relaxation is used as a smoother or preconditioner.}) is given by
\begin{eqnarray}
  C_{\operatorname{horiz}} \approx \frac{\omega^2}{\Delta \xi^2} &\approx& \alpha^2\left(\frac{ c_h\Delta t}{\Delta x}\right)^2.\label{eqn:Choriz}
\end{eqnarray}
As the grid does not have any poles and the ratio between the area of the largest and smallest grid cell is bounded, a typical horizontal grid spacing can be estimated by
\begin{eqnarray}
  \Delta x = \frac{2\pi \Rearth}{4n_x},
\end{eqnarray}
which for $n_x=256$ gives $\Delta x=39$km. On this grid we use a time step of size $\Delta t=10$min, leading to a horizontal coupling of $C_{\operatorname{horiz}} \approx 17.8$. 

Naively, this implies that the constant term in (\ref{eqn:ModelEquation}) is not very important and the equation is very similar to the Laplace or Poisson equation. If, however, a multigrid solver is used to solve the equation, the grid spacing on the coarser multigrid levels increases, which implies that $C_{\operatorname{horiz}}$ decreases. In fact, already after three coarsening steps it is reduced by a factor of $64$ and its magnitude is smaller than one. Thus the coarse grid equation is well conditioned, and even a simple iterative solver, such as SOR or Jacobi will lead to rapid convergence. In a parallel implementation of the multigrid algorithm the ratio between {computation} and communication decreases on the coarser levels and hence using a small number of multigrid levels can improve the parallel scalability. This idea has been explored in the numerical tests reported in Section \ref{sec:ResultsGeoMG}.

Note that in our numerical experiments we adjust $\Delta t$ to keep the Courant number and the ratio $\Delta t/\Delta x$ fixed as the horizontal resolution increases. Hence in these runs $C_{\operatorname{horiz}}$ does not change and for this choice of $\Delta t$ the argument above is independent of the horizontal resolution.
\section{{Iterative} Solvers for elliptic PDEs}\label{sec:SolverReview}
After discretisation 
the linear partial differential equation (\ref{eqn:ModelEquation}) can be written as a sparse matrix equation
\begin{eqnarray}
  A u &=& f. \label{eqn:LinearMatrixEquation}
\end{eqnarray}
The 
vector $u\in\mathbb{R}^n$ represents the discrete solution on the grid such that $u_i$ is the value of the field in {the $i$th} grid cell. Non-linear equations $N[u]=f$ can be solved recursively by a Newton iteration, which {(with a good starting guess)} requires a {(small)} 
number of linear solves.

In the following, several methods for solving the linear equation (\ref{eqn:LinearMatrixEquation}) are discussed and their application in atmospheric models is reviewed. All efficient methods exploit the sparsity of $A$. Some of them, such as geometric multigrid, also use geometric information of the underlying grid. Preconditioners accelerate the speed of convergence by exploiting the structure of the matrix, such as strong vertical coupling. Iterative methods (see e.g. \cite{Freund1992,Templates1994,Golub-vanLoan1996,Saad2003} for a comprehensive treatment) approximate the solution of the equation by a number of iterates $u^{(k)}$, such that {(in exact arithmetic)} $\lim_{k\rightarrow \infty} u^{(k)} = {u}$. 
The most efficient {iterative} solvers only require a small number of iterations $k\ll n$. In any case, due to the presence of discretisation errors and other uncertainties in many meteorological applications it is only necessary to know the solution up to a {fairly large} tolerance.

Most iterative methods do not require the explicit storage of the matrix $A$, it is sufficient to implement the matrix vector operation $y \mapsfrom Ax$.
{The main reason for not explicitly storing the matrix is that on modern computer architectures loading a number from memory is significantly more costly than a floating point operation.}
\subsection{{Preconditioned} Krylov subspace methods}\label{sec:KrylovSpaceMethods}
Krylov subspace methods {iteratively} construct the approximation $u^{(k)}$ in a $k$-dimensional Krylov subspace
\begin{eqnarray}
\mathcal{K}_k=\operatorname{span}\left\{r,Ar,A^2r,\dots,A^{k-1}r\right\} \subset \mathbb{R}^n,
\end{eqnarray}
where $r$ is the initial residual $r=b-Au^{(0)}$.
The simplest {(and in some sense the best)}  Krylov {subspace} method for symmetric positive definite matrices $A$ is the Conjugate Gradient (CG) algorithm by \cite{Hestenes1952}. In every step the approximate solution vector $u^{(k)}$ is updated by adding a vector proportional to the search direction $p^{(k)}$, {such that the energy norm is minimised over $\mathcal{K}_k$}. The search directions are chosen such that they are $A$-orthogonal, i.e. $\langle p^{(k)},Ap^{(k')}\rangle = 0$ for $k\ne k'$. The closely related Conjugate Residual (CR) algorithm is a variant of the algorithm with a different orthogonality constraint. It can be shown (see e.g. \cite{Saad2003}) that the convergence rate depends on the spectral properties of the matrix $A$, in particular on the condition number $\kappa$, which is the ratio between the largest and smallest eigenvalue.
For {finite volume discretisations} of the Poisson equation, $\kappa$ grows rapidly with the inverse grid spacing {$h^{-1}$. It} can be shown that the relative {error} reduction per iteration is $1-2h+\order{h^2}$. Hence the number of iterations required to reduce the error by a factor $\epsilon$ is
\begin{eqnarray}
  k &\propto& \frac{\log \epsilon}{h}.\label{eqn:IterationsCG}
\end{eqnarray}
For anisotropic systems, {such as the one described above}, $h$ is replaced by the smallest grid spacing in the problem, {i.e. $\Delta z$}. Usually the number of iterations can be reduced significantly by preconditioning, as is discussed below. 

As the dominant cost in each step is the matrix application {$y\mapsfrom Ax$}, which is of $\order{n}$ computational complexity, the total cost of the algorithm is \begin{eqnarray}
  \operatorname{Cost}({\operatorname{CG}}) &\propto& \frac{n}{h}\log \epsilon.
  \label{eqn:CostKrylov}
\end{eqnarray}

To solve non-symmetric systems, more general Krylov {subspace} methods such as GMRES, BCG, BiCGStab and GCR can be used {(cf. \cite{Templates1994,Saad2003})}.

As already remarked above, the performance of Krylov {subspace} methods depends on the spectral properties of the matrix $A$. 
{Equivalently, we can multiply the linear system $A\vec{u}=\vec{f}$ by a matrix $M^{-1}$ and solve}
\begin{eqnarray}
  M^{-1}Au &=& M^{-1}f\,.
\end{eqnarray}
This is generally referred to as left preconditioning. $M$ is a matrix with the following properties:
\begin{itemize}
  \item $M$ approximates $A$ well, so that the preconditioned matrix $M^{-1}A$ is {better conditioned than $A$.}
  \item Inversion of $M$ is computationally cheap (again, avoiding explicit storage of $M$).
\end{itemize}
{In general}, these two requirements are mutually exclusive and a tradeoff between them has to be found. Often a good preconditioner can be constructed by using the physical structure of the problem. As discussed in section \ref{sec:VerticalAnisotropy} the operator arising from the discretisation of the pressure correction PDE is highly anisotropic with predominantly vertical couplings, i.e. the relevant grid spacing $h$ in {(\ref{eqn:CostKrylov}) is $\Delta z \ll \Delta x$} and not $\Delta x$. A candidate for the preconditioner would thus be the matrix which only contains the dominant vertical couplings. This matrix is block-diagonal and can be inverted very easily. The result of this is that effectively the number of iterations in (\ref{eqn:IterationsCG}) is set by the horizontal grid spacing {$\Delta x$} instead of the much smaller {$\Delta z$}.

The explicit form of the preconditioned Conjugate Gradient (PCG) algorithm {can be found in \cite{Templates1994,Saad2003}.} 
\ifpreprint
For completeness we have also included it as Algorithm \ref{alg:PCG} in Appendix \ref{sec:Algorithms}. In this form it requires the storage of six vectors ($f,u,z,r,p,q$).
\fi 
At each iteration the following operations have to be carried out:
\begin{itemize}
  \item[1x] {Application of discretised operator} (or matrix-vector product) $y\mapsfrom Ax$
  \item[2x] BLAS level 1 {operations, e.g. $y\mapsfrom a x+y$ (\texttt{axpy})}
  \item[1x] Preconditioner application $y\mapsfrom M^{-1}x$
  \item[{3x}] Scalar products $s\mapsfrom \langle x,y\rangle$
\end{itemize}
Each {application of the operator (and possibly also of the preconditioner)} requires {a local halo-exchange}, and global communication is necessary in the scalar product. The latter usually only accounts for a very small proportion of runtime. 
For other Krylov {subspace} solvers such as GMRES or BiCGStab the number of matrix-vector products, scalar products, and intermediate vectors which need to be stored is different, but the general structure is very similar. {In certain circumstances, when it is safe to carry out a fixed, previously determined number of iterations, the residual norm $||r_K||$ in the stopping criterion is not required and the number of scalar products can be reduced to two.}

\subsection{{Typical Preconditioners}}\label{sec:Preconditioning}
{\em Stationary methods.} Stationary methods were among the first iterative methods to be used in NWP because of their simplicity. For example in \cite{Leslie1973} stationary methods are applied to two dimensional PDEs arising from implicit time stepping in hydrostatic models. A basic overview of the methods discussed below can be found for example in  \cite{Fulton1986}. Even though stationary methods converge slowly on their own, they provide efficient preconditioners for Krylov subspace methods. They are also the main choice for smoothers in multigrid algorithms (see below).

We describe only the successive overrelaxation (SOR) iteration with red-black (RB) ordering of the degrees of freedom in detail, since this method is inherently parallel (in contrast to SOR with lexicographical ordering) and converges faster than the Jacobi iteration.
The grid is split into two sets of `red' and `black' cells, such that the black cells only depend on data in red cells and vice versa, which is always possible for seven point stencils on the lat-long grid or on one panel of the mapped cube grid. For more general grids or discretisation schemes more than two colours may be necessary. By splitting the matrix $A$ into the main diagonal $D$ and upper and lower triangular parts $U$ and $L$, the following two-step recursion can be written down:
\begin{eqnarray*}
  u_r^{(k+1)} &=& (1-\rho)u_r^{(k)}+\rho D^{-1}\left(f_r-(U+L)u_b^{(k)}\right),\\
  u_b^{(k+1)} &=& (1-\rho)u_b^{(k)}+\rho D^{-1}\left(f_b-(U+L)u_r^{(k+1)}\right),
\end{eqnarray*}
where subscripts $r$ and $b$ refer to degrees of freedom associated with red and black cells (resp.).
The overrelaxation parameter $\rho$ can be adjusted to improve convergence.

In parallel implementations, a communication step is necessary after the update of each colour, hence the number of communications is twice that of the Jacobi iteration\footnote{This additional communication can be traded for redundant computations by using a halo which is two grid cells wide.}. On irregular grids with more than two colours more parallel communication may be necessary. This has to be balanced against the better convergence of the SOR method.

Although stationary methods are very easy to implement, they converge in general very slowly, e.g. for the Poisson equation the total computational cost of any pointwise stationary method is
\begin{eqnarray}
  \operatorname{Cost}(\operatorname{stationary}) &\propto& \frac{n}{h^2} \log\epsilon
  \label{eqn:CostStationary}
\end{eqnarray}
as opposed to the $O(h^{-1})$ for Krylov methods in \eqref{eqn:CostKrylov}. 

For anisotropic problem, the convergence can be improved significantly by using block-versions of these algorithms. If the matrix $A$ has a block structure, 
the solution vector can be split into $n_B$ blocks of size $B$,
\begin{eqnarray}
  u &=& \left(\tilde{u}_1,\tilde{u}_2,\dots,\tilde{u}_{n_B}\right)^T.
\end{eqnarray}
In our case, a good blocking is by vertical columns.
The matrix can be written as $A=\tilde{D}+\tilde{U}+\tilde{L}$ with
\begin{eqnarray}
  \tilde{D} &=& \operatorname{diag}\left(\tilde{D}_1,\tilde{D}_2,\dots,\tilde{D}_{n_B},\right),
\end{eqnarray}
where $\tilde{D}_i$ are $B\times B$ block matrices, and with $\tilde{U}$ and $\tilde{L}$ block-upper and block-lower triangular, respectively.
Then for example the first step in the RBSOR iteration (above) can be written for the $i$th red block as
\begin{eqnarray}
  \tilde{u}_{r,i}^{(k+1)} &=& (1-\rho)\tilde{u}_{r,i}^{(k)}\label{eqn:BlockRBSOR}\\&+& \rho\tilde{D}_i^{-1}\left(
  \tilde{f}_{r,i} - \sum_j(\tilde{U}+\tilde{L})_{ij}\tilde{u}_{b,j}^{(k)}
  \right).\notag
\end{eqnarray}
This requires inversion of the \mbox{$B\times B$} matrix $\tilde{D}_i$. In our case, the matrices $\tilde{D}_i$ are tridiagonal and describe the vertical coupling. They can be inverted in $\order{B}$ time with the tridiagonal matrix-algorithm (also known as Thomas algorithm, see e.g. \cite{Press2007}).
\ifpreprint
which is written down explicitly in Algorithm \ref{alg:ThomasAlgorithm} in Appendix \ref{sec:Algorithms}.
\fi 
It is also known as line relaxation because all vertical degrees of freedom in a vertical column are updated simultaneously. For anisotropic problems with predominantly vertical coupling, where $\Delta z \ll \Delta x$, the block version (\ref{eqn:BlockRBSOR}) will converge substantially faster than the point-iteration 
because the cost in \eqref{eqn:CostStationary} is proportional to $\Delta x^{-2}$ and not to $\Delta z^{-2}$.\vspace{2ex}
\noindent
{\em Alternating Direction Implicit method.} The idea of vertical line relaxation can be extended to handle anisotropies in multiple directions or anisotropies that change direction in parts of the domain. These can for example arise due to the convergence of gridlines near the poles on latitude/longitude grids or due to vertical mesh grading. Assuming that we can write $A=A_x+A_y$, where each of the two operators contains only derivatives in one spatial direction (for simplicity considering only two), each iteration of the Alternating Direct Implicit (ADI) method of \cite{Peaceman1955} consists of two steps. In the first step, the term involving $A_x$ is treated implicitly and the term involving $A_y$ is moved to the right hand side. This requires the inversion of the tridiagonal matrix $A_x+\mu I$, where $\mu$ is a parameter. In the second step $A_y$ is treated implicitly and the tridiagonal matrix $A_y+\mu I$ has to be inverted.

Although each step requires only the inversion of a tridiagonal matrix, it is impossible to store $A$ in such a way that both $A_x$ and $A_y$ are tridiagonal simultaneously. It requires reordering some of the field vectors, which reduces cache efficiency and leads to all-to-all communications in a parallel implementation.\vspace{2ex} 
\noindent
{\em Incomplete LU decomposition.} 
An alternative class of preconditioners is based on an approximate factorisation of $A$ into the product of two sparse lower- and upper triangular matrices. Incomplete LU decomposition (ILU) is a modification of the LU decomposition algorithm, which constructs triangular matrices $L$ and $U$ such that $M=LU\approx A$. A description of the algorithm can be found for example in \cite{Saad2003}. The system $LUu=f$ can then be solved in $\order{n}$ time by backsubstitution. In its simplest form with zero fill-in (ILU0), the only modification to the $LU$ factorisation algorithm is that throughout the factorisation process matrix elements $L_{ij}$ and $U_{ij}$ are computed only if $A_{ij} \not= 0$ in the original matrix, i.e. $L$ and $U$ have the same sparsity pattern as $A$. Of course this does not lead to an exact factorisation of $A$. In other versions of the algorithm with higher fill-in the sparsity pattern of $L$ and $U$ is augmented (ILU(p)), or matrix entries are only dropped based on threshold criteria (ILUT).
\subsection{Multigrid methods}
\label{sec:multigridintro}
Stationary methods such as the Jacobi- or SOR iteration reduce the high frequency components of the error first, as they carry out local updates at each grid point. 
After a few iterations, the error is very smooth and the convergence rate deteriorates to the asymptotic value.

The multigrid method (see e.g. \cite{Briggs2000,Trottenberg2001,Hackbusch2003},
a brief overview can also be found in \cite{Fulton1986}) is based on the following insight: whether an error component is classified as being low- or high-frequency depends on the underlying grid; a low frequency error on a fine grid can be interpreted as a high frequency component on a grid of larger grid spacing. By applying the smoother on a hierarchy of coarser grids and using intergrid operators to interpolate data between these levels, all frequency components of the error are reduced simultaneously.

A simple two-grid method works as follows: a small number of iterations of the smoother is applied to reduce the high-frequency components of the error on the fine grid. The smooth residual $r=f-Au$ can be represented well by restriction $r_c=Rr$ on a coarser grid. On the coarse grid the residual equation $A_c e_c = r_c$ is solved. This requires substantially less work as the number of grid points is reduced by a factor of eight (in three dimensions). Finally the coarse grid error is interpolated (prolongated) back to the fine grid $e=Pe_c$ and added to the fine grid solution. Usually a small number of smoothing steps is applied in the end to reduce any high frequency errors introduced by the interpolation.

This two-level method can be extended to a full multilevel method by recursion on a hierarchy of grids with grid spacing $h,2h,4h$, etc. The recursive implementation of a multigrid V-cycle is shown in Algorithm \ref{alg:VCycle}. The fields and system matrix are stored in the arrays $\{u^{(\ell)}\}$, \{$A^{(\ell)}\}$ etc. where $\ell$ is the multigrid level. The finest level is $\ell=L$, whereas the coarsest level corresponds to $\ell=1$. The number of pre- and post-smoothing steps can be specified with $\nu_{pre}$ and $\nu_{post}$.
\begin{algorithm}
\caption{Multigrid V-cycle\newline${}^{}\qquad{}^{}$\texttt{MGVcycle}($\ell,\{A^{(\ell)}\},\{f^{(\ell)}\},\{u^{(\ell)}\},\{r^{(\ell)}\}$)}
\label{alg:VCycle}
\begin{algorithmic}
  \IF{$\ell > 1$}
    \STATE Call \texttt{Smooth}($\ell,\nu_{pre},f^{(\ell)},u^{(\ell)}$) \COMMENT{Presmoothing}
    \STATE $r^{(\ell)} = f^{(\ell)}-A^{(\ell)} u^{(\ell)}$ \COMMENT{Calculate residual}
    \STATE $f^{(\ell-1)} \mapsfrom R_{\ell-1,\ell} r^{(\ell)}$ \COMMENT{Restrict residual}
    \STATE $u^{(\ell-1)} \mapsfrom 0$ \COMMENT{Initialise solution}
    \STATE Call \texttt{MGVcycle}($\ell-1,\{A^{(\ell)}\},\{u^{(\ell)}\},\{f^{(\ell)}\},\{r^{(\ell)}\}$) \COMMENT{Recursion}
    \STATE $e^{(\ell)}\mapsfrom P_{\ell,\ell-1} u^{(\ell-1)}$ \COMMENT{Prolongate solution}
    \STATE $u^{(\ell)} \mapsfrom u^{(\ell)}+e^{(\ell)}$ \COMMENT{Add coarse grid correction}
    \STATE Call \texttt{Smooth}($\ell,\nu_{post},f^{(\ell)},u^{(\ell)}$) \COMMENT{Postsmoothing}
  \ELSE
    \STATE $u^{(1)} = \left(A^{(1)}\right)^{-1} f^{(1)}$ \COMMENT{Solve on coarsest level}
  \ENDIF
\end{algorithmic}
\end{algorithm}

In the three dimensions the cost on each level is usually dominated by the smoother, which has a computational cost of $\order{n}$. With $\nu_{pre}$ and $\nu_{post}$ pre- and  post-smoothing steps, the total cost of one V-cyle is thus approximately
\begin{eqnarray*}
  \operatorname{Cost}(\operatorname{V-cycle}) &\propto& (\nu_{pre}+\nu_{post}) \underbrace{\left(n+\frac{n}{8}+\frac{n}{8^2}+\dots\right)}_{\textstyle \le \frac{8}{7} n} \, ,
\end{eqnarray*}
i.e. the additional computational cost due to the coarser levels is almost negligible. Note, however, that care has to be taken in the parallel implementation as the volume-to-interface ratio decreases on coarser levels, as discussed in \cite{Huelsemann2005}. We will argue in section \ref{sec:GeoMGParallelisation} that this can be avoided for the problem considered in this article without compromising the convergence rate by limiting the number of multigrid levels.

To define the multigrid algorithm, the following components have to be specified by the user:
\begin{itemize}
  \item \textit{Smoother.} This is usually a simple pointwise relaxation method such as Jacobi or SOR, but Incomplete LU factorisation methods can also be used  (see Section \ref{sec:Preconditioning}). However, if the problem has a very strong anisotropy, using line- or plane-relaxation is much more efficient.
  \item \textit{Coarsening strategy.} Usually, the grid spacing is doubled in all dimensions, but other approaches such as semi-coarsening, where only one or two grid dimensions are coarsened, are possible. As discussed in detail in section \ref{sec:TensorProductMultigrid}, horizontal coarsening together with vertical line relaxation is the most robust approach for problems with strong grid aligned anisotropies. In some cases, more aggressive coarsening strategies can be used.
  \item \textit{Intergrid operators.} Various choices exist for the restriction and prolongation operators $R$ and $P$, but it is usually sufficient to use simple methods such as cell-averages for restriction and linear interpolation for prolongation.
  \item \textit{Coarse grid solver.} This can be a direct solver on the coarsest grid, but it doesn't have to be. The coarse grid equation does not need to be solved exactly and so iterative methods can be used to solve the problem approximately. In our case, where the problem is well conditioned on the coarser levels, a small number of iterations of the smoother turns out to be sufficient.
\end{itemize}
It can be shown that the total computational cost to reduce the error by a 
factor $\epsilon$ with multigrid is given by
\begin{eqnarray}
  \operatorname{Cost}(\operatorname{multigrid}) &\propto& n\log\epsilon.
\end{eqnarray}
If the first guess for the solution is constructed starting at the coarsest level (the full multigrid iteration) then multigrid is algorithmically optimal, i.e. the cost to reduce the total error to the size of the discretisation error is proportional to the problem size $n$, independent of the grid spacing. This should be compared to the costs of Krylov subspace and stationary methods in (\ref{eqn:CostKrylov}), (\ref{eqn:CostStationary}).
\subsubsection{Tensor product multigrid}\label{sec:TensorProductMultigrid}
For grid aligned anisotropies it was shown in \cite{BoermHiptmair1999} that a tensor-product multigrid approach with semi-coarsening and line smoothing is most robust and efficient. This is particularly suitable for the strong vertical anisotropy encountered for the elliptic PDEs in NWP, such as the model equation \eqref{eqn:ModelEquation} studied in this article. 

The authors study an operator of the form $-\nabla(\alpha\nabla)$ for a two dimensional model problem. To apply their approach it is necessary that the coefficient $\alpha$ of the underlying operator is separable, so that the resulting discretised matrix $A$ can be written as the sum of two tensor products
\begin{eqnarray}
  A &=& A^{(r)}\otimes M^{(horiz)} \;+\; M^{(r)} \otimes A^{(horiz)},
\end{eqnarray}
where $M^{(\cdot)}$ denotes the mass matrix and $A^{(\cdot)}$ is a second order derivative, either in the radial or horizontal direction. 

They propose to use line relaxation in the radial direction combined with coarsening only in the horizontal direction (semi-coarsening). While this is more costly (but of the same computational complexity) per iteration than the simpler multigrid method with point-relaxation and uniform coarsening, described in the previous section, it can be shown to lead to an optimal multigrid iteration for any grid aligned anisotropy. The convergence rate reduces to the convergence rate for the horizontal operator.

This approach can be extended to three dimensions. If the horizontal problem (which will now be two-dimensional) is isotropic, as is the case for example for the cubed sphere grid employed here, then the grid can be uniformly coarsened in both horizontal directions. This is the method we will use in our numerical tests below. Note however, that the tensor product approach has already been successfully applied to three dimensional problems in NWP by \cite{Buckeridge2010} and \cite{Buckeridge2011} on latitude-longitude grids, where the horizontal coarsening strategy also needs to be suitably adapted.
\subsubsection{Algebraic multigrid}
The geometric multigrid algorithm described so far assumes that the matrix $A$ is based on the discretisation of a PDE on a regular grid and the construction of coarse grid and intergrid transfer operators can be based on geometric information about the underlying grid.

By contrast, algebraic multigrid (see \cite{Brandt1984, Stueben1999}) can be used to solve more general problems of the form $Au=f$ with arbitrary sparse matrices $A$. It is based on the idea that smoothness of the error does not necessarily have to have a geometric meaning, instead the error components which are not reduced by the smoother can be called smooth. It then uses the strength of connection between pairs of points to define a set of coarse grid points, as well as matrix-dependent prolongation and restriction matrices $P$ and $R$. The coarse grid operator is constructed using the Galerkin product $A_c = P^T A R$. In the classical approach the coarse grid consists of a subset of the fine grid points. This should be compared to aggregation based AMG algorithms, where the coarse grid points are obtained by combining several find grid points into coarse grid aggregates.

Algebraic multigrid (AMG) still works best for elliptic PDEs, but it can be applied in circumstances where geometric multigrid has difficulties, for example if the PDE is discretised on an irregular mesh, or if the smoother works only in some directions. Although a theoretical analysis is more difficult, the reasons for the good performance of AMG for heterogeneous and anisotropic elliptic PDEs are fairly well-understood (see \cite{Vassilevski2008} for details). However, it has additional setup costs for the construction of the coarse levels. Also, the matrix $A$ has to be stored explicitly on all levels, whereas for geometric multigrid it can be recomputed at each iteration which can lead to significant efficiency gains (see Section \ref{sec:Bespoke} below). 

Note also that in contrast to geometric multigrid where highly efficient line relaxation can improve convergence, geometric information such as the strong coupling in the vertical direction is not usually exploited directly in AMG, although it is used indirectly in the construction of the coarse grids. Another advantage of the geometric multigrid approach is that the coarse grid operators can be constructed directly by discretisation on the coarse grids, which - for simple discretisations - can lead to a significant smaller stencil than the Galerkin product required for AMG. However, due to the matrix-dependent components, AMG solvers are significantly more robust to coefficient variations of parameters (\cite{Vassilevski2008}).

Both geometric and algebraic multigrid can be used as stand-alone solvers and preconditioners for a Krylov subspace method. However, most AMG solvers, in particular the aggregation based AMG algorithm in \cite{Blatt10}, are not convergent as stand-alone solvers and should always be used as preconditioners for Krylov methods.
\section{Applications in atmospheric modelling}
\subsection{Krylov subspace methods}
Using ADI or stationary methods as preconditioners for Krylov subspace methods has been very popular for solving the pressure correction equation in atmospheric models. 
In \cite{Skamarock1997} a CR solver with ADI preconditioner is used for a non-hydrostatic model. The authors find that on strongly anisotropic grids it is sufficient to only use the vertical part of the ADI iteration. 
\cite{Thomas1997} discuss the use of GMRES(10) with different preconditioners (SOR with lexicographic or red-black ordering and ADI in the vertical direction only) for the MC2 model developed in Canada. The most successful preconditioner (ADI in the vertical direction only) requires $\order{60}$ iterations to reduce the residual by four orders of magnitude for a problem of size $119\times119\times31$, and the number of iterations only increases to around 70 on a $511\times 539\times31$ grid. 
The unpreconditioned GCR algorithm is used in \cite{Smolarkiewicz1994} for a 2d thermal convection problem. The same algorithm is also applied to a semi-Lagrangian Non-Hydrostatic Model in \cite{Smolarkiewiecz1997a} and the authors of \cite{Wu2010} test preconditioners from the PETSc library with a GCR solver. In \cite{Davies05} the dynamical core of a global forecast model is discretised on a latitude-longitude grid. This introduces an additional horizontal anisotropy due to the convergence of gridlines near the poles. For this reason the elliptic PDE in the model is preconditioned with ADI in the vertical and longitudinal direction and solved with a restarted GCR iteration.

Incomplete LU factorisation preconditioners have been used successfully for solving the pressure correction equation in NWP models as well. The authors of \cite{Qaddouri2003} find that a modified ILU(0) preconditioner for GMRES reduced the solution time in the GEM model relative to a direct solver based on Fourier transformations.
In \cite{Zhang2008} the performances of PILUT and Euclid (ILU) solvers from the hypre library were studied as preconditioners for GMRES. The solver was used to solve the pressure correction equation in the GRAPES model. Although in absolute solve times the PILUT preconditioner was not as efficient as BoomerAMG, the number of iterations was comparable for both methods on 24 cores.

A review of iterative methods in meteorological problems can also be found in \cite{Steppeler2003}, which highlights in particular the asymmetry in the Helmholtz equation that terrain following coordinates can cause. As demonstrated in \cite{Skamarock1997}, neglecting these terms can lead to instabilities in the solver.
\subsection{Multigrid}
Multigrid solvers have also already been applied successfully to elliptic PDEs in atmospheric modelling. In \cite{Bates1990}, the multigrid method is used to solve the Helmholtz equation for the geopotential in a shallow water model on the sphere. As described in \cite{Barros1989}, $\lambda$-line relaxation is used to deal with the anisotropy near the poles, but depending on the relative grid spacing in the latitudinal and longitudinal direction, line relaxation in both directions may be necessary in different regions.

\cite{Bowman1991} use a multigrid solver to solve the two-dimensional Helmholtz equation on a sphere. The horizontal anisotropy on the latitude-longitude grid is dealt with by reducing the number of points in the longitudinal direction near the poles on the coarser grids, which makes convergence of the method slightly worse than second order. They find that the time per iteration increases linearly with the number of grid points, as is typical for multigrid methods and a convergence rate per multigrid V-cycle of 0.44, which is worse than the rate that can be achieved on a regular grid. However, by modern standards the finest grid in their numerical studies is relatively coarse with $128\times64$ grid points.

In \cite{Chen2001} the multigrid solvers from the \texttt{mud3cr} package \cite{Adams2001301} are used for solving the Helmholtz equation in a three-dimensional hydrostatic model, as in \cite{Skamarock1997} mixed derivatives are retained in the Helmholtz operator. A set of three realistic model problems with $256\times 256\times 160$ and $512\times512\times320$ grid points are solved. The authors find that to achieve good agreement with the known analytical solution it is sufficient to solve the Helmholtz equation to relatively low accuracy $||r||<0.1$, and that only one or two V-cycles with a point smoother are sufficient to achieve this. For higher accuracies $||r||<0.001$ convergence is achieved in less than 10 iterations, and the use of a line smoother approximately halves the number of iterations (but requires more CPU time). The good performance of the line smoother is reported even for very anisotropic problems with $\Delta z/\Delta x\ll 1$.

A set of Fortran77 subroutines, developed at NCAR, for solving partial differential equations in a two or three dimensional rectangular domain with multigrid methods is described in \cite{Adams1989,Adams1991b} (see also references cited there for further applications). Finite differences are used in the discretisation and the package can handle anisotropies by using both point-, line- and plane smoothers. 

While the problems studied in \cite{Adams1991b} are idealised, the performance of the multigrid solvers in the MudPack package is tested on a set of realistic problems in atmospheric physics in \cite{Adams1992}. In particular a three dimensional model for the (static) flow over orography is studied. On a grid of size $65\times 33\times33$ the multigrid solver, which uses line relaxation in all three spatial directions, converges within 9 iterations, whereas the smoother alone takes more than $3000$ iterations to achieve the same amount of accuracy.
 
Full global models on latitude-longitude grids are discussed in \cite{Buckeridge2010}, where the authors use the tensor-product multigrid approach described above together with selective semi-coarsening near the pole, which can be shown to be optimal. Currently a new dynamical core is under development for the Unified Model \cite{Wood2013}. In its first implementation this uses a BiCGStab solver preconditioned with vertical line SOR, but the implementation of the multigrid method in \cite{Buckeridge2010} is also currently being explored. Preliminary numerical experiments by the authors (in collaboration with the Met Office) on a small operational problem show that, compared to the BiCGStab iteration, using the multigrid solver can reduce the number of iterations significantly and leads to a smaller total solution time. Further tests on larger problems are required to confirm these results.
\subsection{Direct solvers and spectral methods}
Direct methods can usually only be applied for relatively small systems or on grids that have a particular structure and size. However, some more advanced direct methods, such as block methods and cyclic reduction are discussed in \cite{Leslie1973} and applied to small two dimensional problems, arising for example from semi-implicit discretisations in hydrostatic models. The authors compare direct and iterative methods for solving problems of the Poisson- and Helmholtz type. 

Alternatively, if the operator has a tensor-product structure, one can use eigenmode expansions. This is done for two-dimensional problems, which arise from a vertical mode decomposition in \cite{QaddouriLee} and \cite{Qaddouri2003} where the performance of direct and iterative solvers in the Canadian Limited Area Forecasting Model GEM-LAM are compared. Expanding the right hand side in the eigenmodes in one direction results in a set of tridiagonal systems (one for each mode) of size $n'$, which can be solved by the Thomas algorithm. On regular grids with a suitable number of of points, one can use fast-Fourier transformations for the projection on eigenmodes, which have a cost that grows with $O(n \log n)$. 

Not surprisingly, \cite{Qaddouri2003} find that for larger problem sizes the performance of the iterative conjugate gradient solver preconditioned with variants of incomplete LU outperforms the direct solver unless fast Fourier decomposition can be used. However, \cite{Hess1997} demonstrate that the fast Fourier transform method is not as fast as a multigrid algorithm. The authors stress that the small timestep and resulting large constant term in the Helmholtz equation will improve the convergence of the iterative method, but has no impact on the performance of the direct method, similar observations are reported in \cite{Leslie1973}.
One of the other disadvantages of direct methods is that they are hard to parallelise as they require global communications in the Fourier transformation. 

Spectral methods are also used by the forecast model of the European Centre of Medium Range Weather Forecasts (ECMWF). Recently, a fast Legendre transformation methods has been implemented which improves the performance and scalability of the model (see \cite{Wedi2013}).
\subsection{Scalability}
The parallel performance of the multigrid solver in \cite{Adams1991b} is demonstrated for a set of test problems, including a three dimensional Helmholtz problem on (part of) a latitude/longitude spherical grid ($\pi/4\le\theta\le3/4\pi$; thus avoiding the pole problem). These tests were carried out with up to $193\times65 \times129$ grid points on a CRAY Y-MP8/864 vector machine. Solving up to an error of $0.25\cdot 10^{-4}$ (discretisation error) takes 1.91s, but there is an additional overhead of 51.61s for initialisation (which includes discretisation and (tridiagonal) matrix factorisation). Line relaxation is used in the radial direction for this problem. 

The multigrid solver studied in \cite{Hess1997} shows very good strong scalability with parallel efficiencies exceeding $70\%$ on up to 32 processors on the CM-5 and Cenju-3 systems used in the study; a 2d problem of size $1025\times1025$ was solved. On a problem of this size the iterative solver outperforms the FFT-based direct solver due to its lower complexity of $\order{n}$ instead of $\order{n\log n}$.

In \cite{Thomas1997} strong scaling tests for the entire dynamical core (including the solver, but also other model components) have been carried out both on a Cray-TE3 system and NEC SX-4 supercomputer. For a local problem size of at least $321\times321\times31$ points the model shows good scaling up to 70 cores on the Cray-TE3 machine: the performance is around 30 MFlops/s per processing unit (5\% of the theoretical peak performance). Problem of sizes up to $502\times1936\times10$ were run on one node of the NEC SX-4, and here it was found that the performance only drops by around $10\%$ when going from one to 32 processes.

Scaling tests with a 3d multigrid solver on the Cray Jaguar XT5 machine have been presented in \cite{Nyberg2010}. The solver is part of a global cloud resolving model on a geodesic grid, developed by David Randall at Colorado State University. The multigrid solver scaled to 80,000 cores on a Jaguar XT5 machine; the largest considered system had $8.6\cdot10^{10}$ degrees of freedom and was solved with 20 V-cycles in $17.166s$ on 81,920 cores. Strong scaling from 20,480 to 81,920 cores was good for the same system.

In \cite{Zhang2008} various preconditioners for the Helmholtz equation encountered in the GRAPES non-hydrostatic local area model are investigated.
The PETSc environment is used with preconditioners from the hypre library. The four preconditioners that were tested are BoomerAMG, SAI (Parasails), PILUT and Euclid (ILU), in addition to the Jacobi preconditioner in PETSc. The solvers were used to solve a (sign-positive) Helmholtz problem from a regional scenario of size $37\times31\times17$. The best serial performance is achieved with the BoomerAMG preconditioner, which requires 9 iterations to convergence. On 16 cores the number of iterations for the BoomerAMG preconditioner doubles relative to the sequential run. In contrast, the number of iterations for the PILUT preconditioner decreases significantly on larger core counts and is less than that for BoomerAMG on 24 cores. In terms of absolute times Parasail, which has a constant number of iterations, independent of the number of cores, gives the best performance on 16 cores.

The solver for the Helmholtz equation in the GRAPES-global model is discussed in \cite{Wu2010}. The three dimensional equation is solved with a preconditioned GCR solver as well as with a Krylov subspace solver in PETSc with a hypre preconditioner. The GCR preconditioner is constructed by only retaining the largest elements of the discretisation matrix. Strong scaling tests from 64 to 256 processors are shown, and the solution time could be reduced by using PETSc.

The authors of \cite{QaddouriLee} and \cite{Qaddouri2003} study the parallel scaling of two direct methods and an iterative solver (GMRES, preconditioned with the direct solver on each domain) on up to 1600 cores. They find that the slow Fourier transformation method scales very poorly and is significantly slower than the iterative method. The runs are carried out on an IBM p575+ cluster with 121 compute nodes, where each node contains 16 processors. It is found that while the FFT solver outperforms the iterative solver in absolute times, the latter shows better scalability. For the largest problem size the iterative solver takes $195.85$s to solve a problem with $1.2\cdot10^{9}$ degrees of freedom. The number of iterations is stable and does not exceed 5. However, the FFT solver can only be used for certain grid sizes and the authors find that if the ``slow'' Fourier transformation is used, the direct solver is outperformed by the iterative solver on larger problem sizes. A series of weak scaling tests with problem sizes of up to $3852\times3852$ horizontal degrees of freedom and 80 vertical levels on up to 1600 processors are presented.
\section{Implementation}\label{sec:Implementation}
In this work we implemented a range of algorithms for solving the model equation (\ref{eqn:ModelEquation}) on one panel of a gnomonic cubed sphere grid (but see also Section \ref{sec:ResultsDUNEGrid} which describes the performance of a bespoke geometric multigrid solver on the entire sphere). 
To evaluate the performance of existing solver packages we tested and optimised the algebraic multigrid solvers in the DUNE and hypre libraries.
In addition we implemented a matrix-free Conjugate Gradient solver, which uses vertical line relaxation as a preconditioner and wrote a bespoke geometric multigrid algorithm following the tensor product idea in \cite{BoermHiptmair1999}. 
\subsection{AMG solvers in DUNE and hypre}
The Distributed and Unified Numerics Environment (DUNE) is a modular C++ library for the solution of PDEs with grid based methods. The DUNE-Grid library \cite{Bastian2008a,Bastian2008b} provides interfaces to various parallel grid implementations such as ALUGrid \cite{Dedner2004,Burri2005}, but also implements its own grids. To implement the cubed sphere grid, we  used the \texttt{GeometryGrid} class, which describes a mapping from a simple unit cube to curved coordinates.
Several discretisation packages, such as DUNE-PDELab can be used to translate a local operator into a mapping on a grid function space and finally into a sparse matrix in compressed sparse row storage (CSR) format. The Iterative Solver Template Library (ISTL) \cite{Blatt07,Blatt08} provides (parallel) solvers for solving the sparse matrix equation $Au=f$, including various Krylov subspace methods such as Conjugate Gradient and BiCGStab, as well as basic iterative methods such as Jacobi or SOR, and preconditioners such as ILU0. In particular, it includes an aggregation-based parallel algebraic multigrid algorithm, described in \cite{Blatt10}.
\subsubsection{Optimisation}
The default parameter settings in the ISTL AMG solver are for isotropic problems and had to be adapted for our case. We varied the parameters \texttt{maxDistance} (default: 4), which controls the maximal distance between points in an aggregate, and \texttt{prolDampFactor} (default: 1.6) which is the factor by which the coarse grid correction is multiplied before it is added to the fine grid solution. In general, the time per iteration scaled very well for any parameter setting, but the number of iterations could be reduced significantly by changing the two parameters mentioned above. The optimal value for \texttt{maxDistance} turned out to be 3. As can be seen in Fig. \ref{fig:AMGIterationsprolDampFactor}, the number of iterations is very sensitive to \texttt{prolDampFactor}, and we found that the optimal value is actually 1.0. 
\begin{figure}
  \begin{center}
  \includegraphics[angle=270,width=\linewidth]{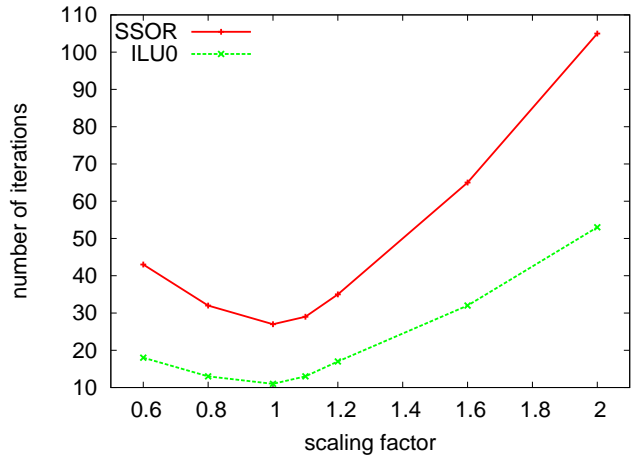}
  \caption{Number of iterations of the DUNE ISTL AMG solver for different values of the prolongation damping factor and two smoothers. The runs were carried out with 64 cores on a $512\times 512\times 128$ grid.}
  \label{fig:AMGIterationsprolDampFactor}
  \end{center}
\end{figure}

We also implemented an interface to the BoomerAMG solver from the hypre library \cite{FalgoutYang2002,Falgout2006} within DUNE.
Again the default BoomerAMG parameters had to be adapted for our problem, in particular for large processor counts. In general, this improved the scalability of the code at the cost of a small increase in the number of iterations. We found that the following parameters gave the best results (see the hypre reference manual for more detailed explanations of the parameters):
\begin{itemize}
  \item Algebraic coarsening strategy (\texttt{coarsentype}): HMIS coarsening (default: Falgout coarsening)
  \item Maximal number of matrix elements per row (\texttt{pmaxelmts}): 4 (default: 0, i.e. not limited)
  \item Number of aggressive coarsening steps (\texttt{aggnumlevels}): 2 (default: 0)
\end{itemize}

In both cases the AMG solvers were used as preconditioners for a  Conjugate Gradient solver. To use BoomerAMG as a preconditioner within CG, the order of relaxation (\texttt{relaxorder}) had to be changed from the default (CF ordering) to lexicographic ordering. Finally, we also used ILU0 as a preconditioner for the CG solver.
\subsection{Bespoke matrix-free solvers}
\label{sec:Bespoke}
\subsubsection{Geometric multigrid}\label{sec:ImplementationGeometricMultigrid}
Although they can be applied in very general circumstances, a small draw-back with algebraic multigrid solvers is the setup costs associated with the construction of the multigrid hierarchy and the larger storage requirements due to the explicit storage of the matrix $A$, as well as of the coarse grid hierarchy. To avoid this we implemented a matrix-free geometric multigrid code based on the tensor product idea 
described in section \ref{sec:TensorProductMultigrid}. Line relaxation in the vertical direction is combined with semicoarsening in the horizontal direction. 
For simplicity the code was implemented on a regular three dimensional grid with $n=n_x\times n_x \times n_z$ grid cells which represents the atmosphere on one panel (i.e. 1/6th) of the surface of a cubed sphere grid. The mapping from the unit square to the surface of the sphere is implemented by including the appropriate geometric factors in the matrix stencil.

In the numerical tests we used a block-RB SOR smoother as described in Section \ref{sec:Preconditioning}. The blocks correspond to data in vertical columns and columns are labelled as red (black) if the sum of their horizontal indices is even (odd). Data was restricted to coarser levels using a simple cell average in the horizontal direction and linear interpolation was used for prolongation to finer grids. In most cases one iteration of the smoother was used to solve the coarse grid equation (see Section \ref{sec:Robustness}).

For cache efficiency it is essential that data in a vertical column is stored consecutively in memory. This can be achieved by storing three dimensional fields $u_{ijk}$ lexicographically in an array $U$ of length $n$, such that
\begin{eqnarray}
  U_m &=& u_{ijk} \quad\text{with}\quad m=n_z (n_x \cdot i+j) + k
\end{eqnarray}
where $(i,j)$ are the horizontal indices and $k$ is the vertical index.
Then for each $(i,j)$ the solution of the tridiagonal system in the block smoother only requires operating on the consecutive data $U_{n_z(n_x\cdot i+j)+1},\dots,U_{n_z(n_x\cdot i+j)+n_z}$. Note that we never store the matrix entries explicitly, they are recalculated whenever they are used in the algorithm. In particular we exploit the tensor product structure of the operator as already described in \cite{Mueller2013a}: while the vertical derivative and mass matrices, which do not vary from column to column and can hence be kept in cache, are stored explicitly, the horizontal discretisation is recalculated from the geometric factors. The latter has to be done only once per column and will hence lead to a very small overhead.

We also implemented the same code in the DUNE-grid interface, based on an arbitrary two dimensional grid and demonstrated that essentially the same performance can be achieved as in the hand-written Fortran code if the number of grid cells in the vertical direction is large enough to ``hide'' the overhead of indirect addressing in the horizontal direction as suggested in \cite{MacDonald2011}. The details of this implementation will be described in a forthcoming publication \cite{Dedner2012}.

As already mentioned in Section \ref{sec:HorizontalCoupling}, the Helmholtz equation is better conditioned on the coarser multigrid levels. In particular, the relative strength of the horizontal coupling, i.e. the size of the off-diagonal matrix entries, is given on level $\ell$ by 
\begin{eqnarray}
  C_{\operatorname{horiz}}(\ell) = C_{\operatorname{horiz}} \times 2^{-2\ell}
\end{eqnarray}
with\vspace{-1ex}
\begin{eqnarray}
  C_{\operatorname{horiz}} &=&\alpha^2 \left(\frac{c_h \Delta t}{\Delta x}\right)^2 \approx 17.8 \approx 2^{2 \cdot 2.077}
\end{eqnarray}
as in (\ref{eqn:Choriz}). Hence already on the third coarse multigrid level ($\ell=3$) the matrix is very well conditioned and line relaxation will be very efficient as a stand-alone solver. Additional multigrid levels will not improve the convergence of the V-cycle significantly.

The robustness of the algorithm with respect to the number of multigrid levels is studied in Section \ref{sec:Robustness}.
\subsubsection{Conjugate Gradient}
To compare to the performance of a typical one-level method, we also implemented a Conjugate Gradient solver preconditioned with vertical line relaxation (block RB SOR). As for the tensor-product multigrid solver, the matrix is not stored explicitly and the matrix elements are recalculated whenever they are needed. Note that in addition to the local halo exchanges in the multigrid algorithm the CG solver requires global communication to evaluate global sums due to dot products.
\subsubsection{Parallelisation}\label{sec:GeoMGParallelisation}
In all cases the domain was partitioned in the horizontal direction only, as is common in atmospheric models; the number of processor is $2^{2p}$. The horizontal partitioning implies that vertical columns are always kept on one processor, which facilitates the use of the Thomas algorithm for the inversion of the tridiagonal matrices in the line relaxation. \cite{Piotrowski2011} have applied parallel tridiagonal solvers to atmospheric models before, but we do not pursue this any further here.

Details on the parallelisation of the DUNE and hypre libraries can be found in \cite{Bastian2008a,Bastian2008b,Blatt08,Blatt10} and in \cite{FalgoutYang2002,Falgout2006}, respectively.

Parallelisation of the matrix-free Conjugate Gradient solver is straightforward: a halo exchange is required after each smoothing step in the preconditioner application and global sums need to be evaluated with the appropriate \verb!MPI_reduce()! calls.

To parallelise the multigrid V-cycle in Algorithm \ref{alg:VCycle} in Section \ref{sec:multigridintro} we proceed as follow: assuming that on entry the solution vector $u$ is consistent on the halo cells (i.e. the entries in each physical grid cell agree on neighbouring processors), we have to add
\begin{eqnarray}
  N_{halo} &=& 1 + s (\nu_{pre}+\nu_{post})
\end{eqnarray}
halo exchanges on each level, where $s=2$ for RB ordering and $s=1$ otherwise, i.e. one halo exchange after each relaxation step (two in the case of RB ordering) and one after the prolongation. Typically we use one pre- and post-smoothing step for RB Line SOR relaxation, resulting in 5 halo exchanges on each multigrid level. The code was optimised by overlapping calculations and communications for the halo exchanges. To do this, the columns at the boundary of the domain were relaxed first and an asynchronous send and receive was posted for the halo data before relaxing the interior columns. The same mechanism was used for the prolongation operation.

On the coarser levels the number of horizontal columns can be smaller than the number of processors. Then data is only stored on a subset of processors of size $2^{2q}$, with $0 \le q < p$. All other processors are idle, see Fig. \ref{fig:ParallelCollectDistribute}. In addition to the total number of levels $L$ we define a level $L_{split}$, where we start reducing the number of processors by pulling together data with the \verb!Collect()! subroutine (the opposite operation is \verb!Distribute()!). This reduction will then be done on every successive level until all data is stored on one processor or until the coarsest level is reached, see Figs. \ref{fig:ParallelCollectDistribute} and \ref{fig:ParallelMGVCycle}.
\begin{figure}
\begin{center}
  \includegraphics[width=0.6\linewidth]{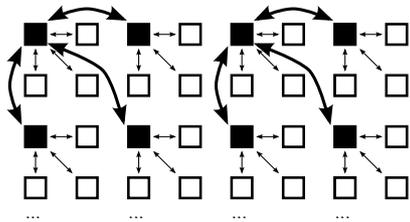}
\caption{Parallel collect and distribute operations. All processors store data for $\ell=L$. Only the gray and black processors store data at level $\ell=L-1$ and only the black processors store data at level $L-2$. Collect and distribute operations between processors on level $L-1$ are indicated by thin arrows, operations on level $L-2$ are shown by thick arrows.}
\label{fig:ParallelCollectDistribute}
\end{center}
\end{figure}

\begin{figure}
  \begin{center}
    \includegraphics[width=0.9\linewidth]{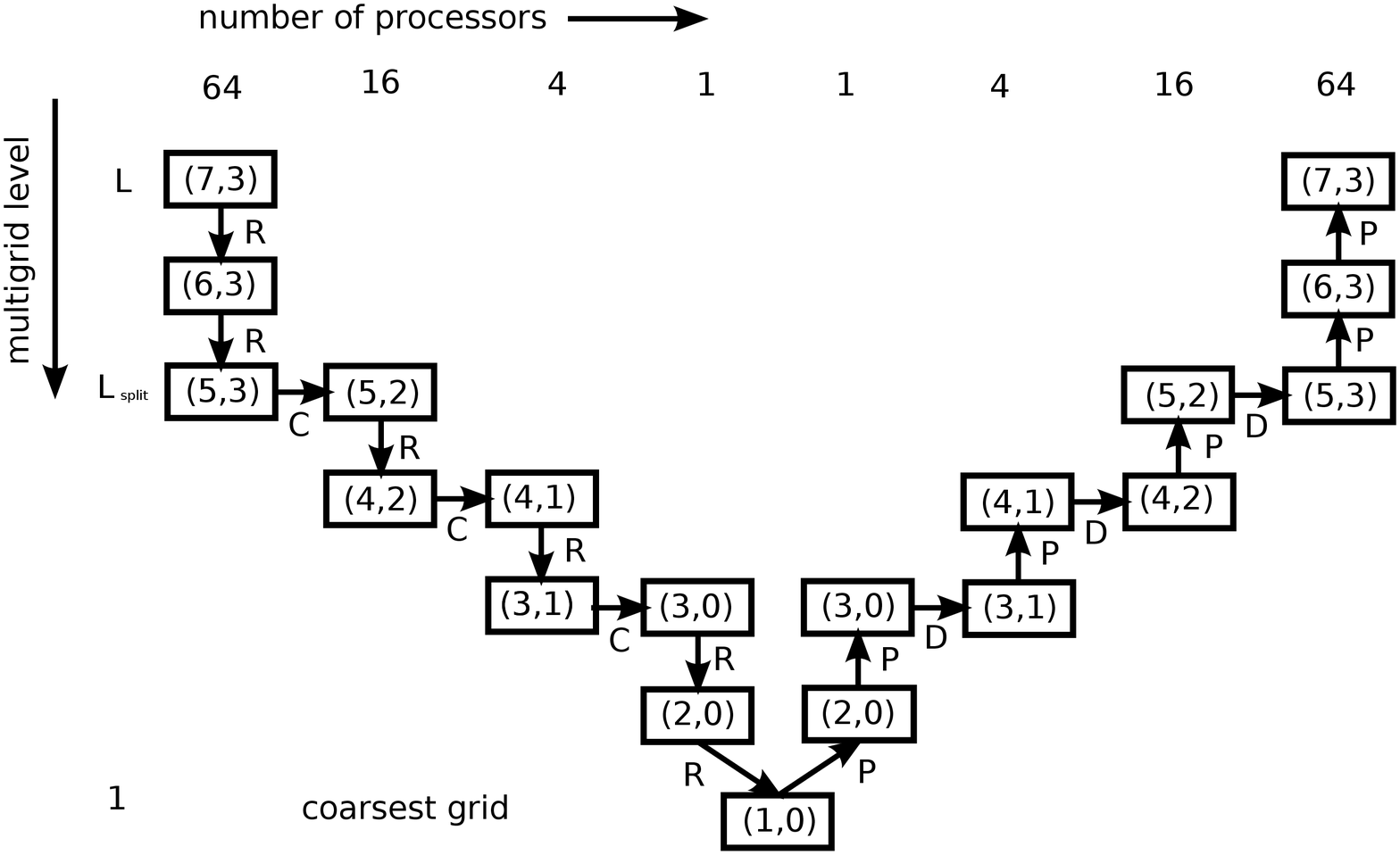}
    \caption{Parallel multigrid V-cycle. The multigrid level decreases from top to bottom. The number of processors is shown along the horizontal axis. Each box symbolises a grid denoted by $(\ell,q)$, defining the multigrid level and the number of processors $2^{2q}$. The arrows indicate restriction (R), prolongation (P), collect (C) and distribute (D). In this example, the coarsest grid is set up on one processor.}
  \label{fig:ParallelMGVCycle}
  \end{center}
\end{figure}
As the problem we solve is very well conditioned on the coarser levels, it might be sufficient to only use a small number of multigrid levels. If we only coarsen until one or more columns per processor are left, it is not necessary to collect and distribute data on the coarser levels. As demonstrated in our numerical experiments below, this does indeed help to improve the scalability of the algorithm without any negative impact on the convergence rate.
\section{Numerical results}\label{sec:Results}
In the following we demonstrate the scalability and robustness of the solvers described above. All runs were carried out on the Phase 3 configuration of the HECToR supercomputer (see \texttt{www.hector.ac.uk}), which consists of 2816 compute nodes. Each node contains two 16-core AMD Opteron 2.3GHz Interlagos chips; in total this amounts to 90,112 cores. Each 16 core processor shares 16GB of memory, which amounts to 1GB per core. The nodes are connected via a Cray Gemini interconnect and organised in a 3D torus. The MPI point-to-point bandwidth is quoted as 5GB/s and the latency between two nodes as around $1.0-1.5\mu s$. The code was compiled with the gnu c++/gfortran compiler and the MPICH2 library.
\subsection{Weak scaling assumptions}
All runs were carried out on one panel of the cubed sphere grid with the vertical grid size kept fixed at \mbox{$n_z=128$} (but the runs were also repeated with $n_z=256$). In the weak scaling runs the number of grid cells per processor is kept constant, the total number of grid cells increases up to $3.4\cdot10^{10}$ on the finest grid, which is run on 65536 processors.

As the number of cells $n_x$ in one horizontal direction is increased, the physical grid spacing \mbox{$\Delta x=\Delta y\propto 1/n_x$} decreases. To keep the acoustic Courant number $c_h\Delta t/\Delta x\approx 8.4$ fixed, the time step size $\Delta t$ is reduced accordingly on finer grids. Note that, as discussed above, the size of the horizontal couplings relative to the zero order term remains constant at around $17.8$.
The ratio $\lambda^2$ tends to the limiting value of 1 as $\Delta t\rightarrow 0$. The parameter space is shown in Tab.~\ref{tab:ParameterSpace}, where we also list the anisotropy $\beta=\lambda^2(\Delta x/\Delta z)^2$ at the bottom, middle ($k=n_z/2$) and top of the atmosphere.
\begin{table*}
  \renewcommand{\arraystretch}{1.25}
  \begin{center}
  \begin{tabular}{|cr|rrrrrrr|}
  \hline
  $n_x\times n_x\times n_z$ & \# dof & 
    $\Delta x$ [km] & $\Delta t$ [s] & $\omega^2$ & $\lambda^2$ &
    $\beta_{\operatorname{bottom}}$ & 
    $\beta_{\operatorname{middle}}$ & 
    $\beta_{\operatorname{top}}$ \\\hline\hline
  $256\times256\times128$ & $8.3\cdot10^6$ & 
    $ 39.1$ & $600.0$ & $6.71\cdot10^{-4}$ & $3.32\cdot10^{-2}$ &
    $3.35\cdot 10^{6}$ & $201.4$ & $51.5$ \\
  $512\times512\times128$ & $3.4\cdot10^7$ & 
    $ 19.5$ & $300.0$ & $1.68\cdot10^{-4}$ & $1.21\cdot10^{-1}$ &
    $3.05\cdot10^{6}$ & $183.1$ & $4.69$ \\
  $1024\times1024\times128$ & $1.3\cdot10^8$ &
    $  9.8$ & $150.0$ & $4.19\cdot10^{-5}$ & $3.54\cdot10^{-1}$ &
    $2.24\cdot10^{6}$ & $134.5$ & $34.4$ \\
  $2048\times2048\times128$ & $5.4\cdot10^8$ & 
    $  4.9$ & $ 75.0$ & $1.05\cdot10^{-5}$ & $6.87\cdot10^{-1}$ &
     $1.09\cdot 10^{6}$ & $65.2$ & $16.7$\\
  $4096\times4096\times128$ & $2.1\cdot10^9$ & 
    $  2.4$ & $ 37.5$ & $2.62\cdot10^{-6}$ & $8.98\cdot10^{-1}$ & 
   $3.54\cdot10^{5}$ & $21.3$ & $5.45$ \\
  $8192\times8192\times128$ & $8.6\cdot10^9$ &
    $  1.2$ & $ 18.8$ & $6.55\cdot10^{-7}$ & $9.72\cdot10^{-1}$ &
    $9.60\cdot10^{4}$ & $5.77$ & $1.48$ \\
  $16384\times16384\times128$ & $3.4\cdot10^{10}$ & 
    $  0.6$ & $  9.4$ & $1.64\cdot10^{-7}$ & $9.93\cdot10^{-1}$ &
    $2.45\cdot10^{4}$ & $1.47$ & $0.38$ \\
  \hline
  \end{tabular}\\[1ex]
  \caption{Parameter space. The last three columns show the anisotropy $\beta=\lambda^2(\Delta x/\Delta z)^2$ at the bottom, middle
and top of the atmosphere.}
  \label{tab:ParameterSpace}
  \end{center}
\end{table*}
The horizontal grid is subdivided into $\mathcal{P} = 2^{2p}$ subdomains} of equal size, each assigned to one processor. Weak scaling then amounts to a fourfold increase of the number of processors whenever the horizontal grid resolution is doubled.

In all cases we initialised the solution with zero and iterated until the residual was reduced by a factor of $10^{-5}$.
\subsection{Preconditioned Krylov subspace methods}
We used two different implementations of the Conjugate Gradient (CG) algorithm with two different preconditioners: (i) the CG solver in the DUNE-ISTL framework with ILU0 preconditioner, where the matrix was set up with DUNE-PDELab (introducing additional matrix setup costs), and (ii) a separate Fortran code which avoids explicit storage of the matrix and uses a vertical line relaxation preconditioner (RB block SSOR). As the matrix stencil is recalculated `on-the-fly' whenever it is needed, there are no additional matrix setup costs in the second case.

In Tab. \ref{tab:WeakScalingOneLevel} the number of iterations, time per iteration and total solution time for both methods are shown as a function of the problem size $n$ and of the number of cores $\mathcal{P}$. In addition, we calculate the scaled parallel efficiency for the time per iteration (relative to a run on $\mathcal{P}_0$ cores) as follows:
\begin{eqnarray}
  E_S(\mathcal{P}) &=& \frac{\titer(\mathcal{P}_0;n=\mathcal{P}_0 * n_\text{loc})}{\titer(\mathcal{P}; n=\mathcal{P} * n_\text{loc})}
\end{eqnarray}
Here, $n_\text{loc}$ is the number of degrees of freedom per subdomain/processor.

The two solvers perform similarly in terms of the number of iterations, which is expected due to the strong vertical coupling. However, the parallel scaling of the ILU0 preconditioner deteriorates beyond 1024 cores and then stays roughly constant. We have no explanation for this.

While the numerical results presented here are for a Conjugate Gradient solver, which can only be applied to symmetric systems, we expect similar results for other more general Krylov subspace methods as these contain the same type of operations in the inner loop. For example, the BiCGStab algorithm requires exactly twice the number of sparse matrix-vector products, preconditioner solves, \texttt{axpy} operations and global reductions as the Conjugate Gradient algorithm.
\begin{table*}
  \renewcommand{\arraystretch}{1.25}
 \begin{center}
 \begin{tabular}{|rr|rrrr|rrrr|}
   \hline
    & &
      \multicolumn{4}{|c|}{CG + ILU0} & 
      \multicolumn{4}{|c|}{CG + line relaxation} \\
   \# cores ($\mathcal{P}$) & \# dof & 
      \# iter & $\titer$ & $E_S(\mathcal{P})$ & $\tsolve$ &
      \# iter & $\titer$ & $E_S(\mathcal{P})$ & $\tsolve$ \\
   \hline\hline
   16 & $8.3\cdot10^{6}$ & 
     74 & 0.235 & --- & 17.41 & 
     44 & 0.109  & --- & 4.78 \\
   64 & $3.4\cdot10^{7}$ & 
     71 & 0.273 & $86\%$ & 19.37 & 
     43 & 0.113 & $96\%$ & 4.88 \\
   256 & $1.3\cdot10^{8}$ & 
     67 & 0.774 & $30\%$ & 51.86 & 
     41 & 0.114 & $96\%$ & 4.66\\
   1024 & $5.4\cdot10^{8}$ & 
     54 & 1.272 & $18\%$ & 68.68 & 
     41 & 0.116 & $94\%$ & 4.75\\ 
   4096 & $2.1\cdot10^{9}$ & 
     56 & 1.419 & $17\%$ & 79.47 &
     41 & 0.117 & $93\%$ & 4.81 \\
   16384 & $8.6\cdot10^{9}$ & 
     50 & 1.382 & $17\%$ & 69.12 & 
     40 & 0.115 & $94\%$ & 4.73\\
   65536 & $3.4\cdot10^{10}$ & 
        &      &  & & 
     40 & 0.115 & $94\%$ & 4.73\\
   \hline  
  \end{tabular}\\[1ex]
  \caption{Weak scaling results for the one-level method. Number of iterations, time per iteration and scaled parallel efficiency $E_S(\mathcal{P})$ for two different preconditioned conjugate gradient implementations with $n_\text{loc} = 2^{19}$ and $\mathcal{P}_0=16$. All times are given in seconds.}\label{tab:WeakScalingOneLevel}
 \end{center}
\end{table*}
\subsection{Multigrid methods}
The number of iterations can be reduced by using multilevel methods.
Results for the weak scaling of the geometric and algebraic multigrid solvers are shown in Tab. \ref{tab:MultigridTimePerIteration} and the total solution times are listed in Tab. \ref{tab:MultigridTotalTimes}  (split up into solve time and coarse grid setup time for the AMG preconditioners). For the DUNE AMG solver ILU0 was used as a smoother (we also carried out runs with an SSOR smoother, but that lead to an increase in the number of iterations by roughly a factor of 2), whereas SSOR relaxation was used in BoomerAMG. Both AMG solvers were used as preconditioners for the DUNE-ISTL Conjugate Gradient algorithm; the geometric multigrid code was used as a standalone solver. In addition, the two AMG preconditioners also require the setup of the matrix~$A$. However, we did not quantify or include this here, since DUNE-PDELab is not optimised for our simple finite volume discretisation and thus the matrix setup time would be vastly overestimated. It is an additional factor in favour of the geometric multigrid code though.
\begin{table*}
\renewcommand{\arraystretch}{1.25}
\begin{center}
 \begin{tabular}{|rr|rrr|rrr|rrr|}
   \hline
    &  & 
    \multicolumn{3}{c|}{AMG (DUNE)} & 
    \multicolumn{3}{c|}{BoomerAMG (hypre)} &
    \multicolumn{3}{c|}{geometric MG} \\
   \# cores ($\mathcal{P}$) & \# dof & 
    \# iter & \titer & $E_S(\mathcal{P})$ & 
    \# iter & \titer & $E_S(\mathcal{P})$ & 
    \# iter & \titer & $E_S(\mathcal{P})$ \\
   \hline\hline
   16 & $8.3\cdot10^{6}$ & 
      12 & 0.56 & --- & 
      12 & 0.73 & --- &
      6 & 0.143 & --- \\
   64 & $3.4\cdot10^{7}$ & 
      13 & 0.56 & 100\% & 
      13 & 0.73 & 100\% &
      6 & 0.148 & 97\% \\
   256 & $1.3\cdot10^{8}$ & 
      12 & 0.59 & 96\% & 
      12 & 0.75 & 97\% &
      6 & 0.152 & 94\% \\
   1024 & $5.4\cdot10^{8}$ & 
      14 & 0.60 & 95\% & 
      12 & 0.75 & 97\% &
      6 & 0.155 & 93\% \\
   4096 & $2.1\cdot10^{9}$ & 
      14 & 0.59 & 96\% & 
      12 & 0.75 & 97\% &
      6 & 0.159 & 90\% \\
   16384 & $8.6\cdot10^{9}$ & 
      14 & 0.60 & 94\% & 
      11 & 0.86 & 84\% &
      6 & 0.161 & 89\% \\
   65536 & $3.4\cdot10^{10}$ & 
      11 & 0.62 & 91\% & 
      9 & 2.24 & 32\% &
      6 & 0.177 & 81\% \\
   \hline
 \end{tabular}\\[1ex]
  \caption{Number of iterations, time per iteration and scaled parallel efficiency for different multigrid solvers ($n_\text{loc} = 2^{19}$, $\mathcal{P}_0=16$). The AMG algorithms were used as preconditioners for CG, whereas the geometric multigrid algorithm was used as a stand-alone solver. All times are given in seconds.}
  \label{tab:MultigridTimePerIteration}
\end{center}
\end{table*}
\begin{table*}
  \renewcommand{\arraystretch}{1.25}
 \begin{center}
 \begin{tabular}{|rr|rrrrr|rrrrr|r|}
   \hline
    & &
      \multicolumn{5}{|c|}{AMG (DUNE)} & 
      \multicolumn{5}{|c|}{BoomerAMG (hypre)} & 
      geometric MG \\
   \# cores & \# dof & 
      \tsolve & + & \tsetup & = & \ttotal &
      \tsolve & + & \tsetup & = & \ttotal &
      \ttotal \\
   \hline\hline
   16 & $8.3\cdot10^{6}$ & 
    6.78 & + & 3.26 & = & 10.04 & 
    8.72 & + & 2.59 & = & 11.31 & 
    0.86 \\
   64 & $3.4\cdot10^{7}$ & 
     7.30 & + & 3.80 & = & 11.10 & 
     9.52 & + & 2.74 & = & 12.26 &
     0.89 \\
   256 & $1.3\cdot10^{8}$ & 
     7.02 & + & 4.53 & = & 11.55 & 
     8.98 & + & 2.82 & = & 11.80 & 
     0.91 \\
   1024 & $5.4\cdot10^{8}$ & 
     8.36 & + & 4.92 & = & 13.28 & 
     9.04 & + & 3.18 & = & 12.22 & 
     0.91 \\ 
   4096 & $2.1\cdot10^{9}$ & 
     8.23 & + & 5.00 & = & 13.23 & 
     8.99 & + & 3.56 & = & 12.55 &
     0.93 \\
   16384 & $8.6\cdot10^{9}$ & 
     8.44 & + & 6.32 & = & 14.76 & 
     9.43 & + & 5.75 & = & 15.18 & 
     0.95 \\
   65536 & $3.4\cdot10^{10}$ & 
     6.80 & + & 9.76 & = & 16.56 & 
     20.20 & + & 7.09 & = & 27.29 & 
     1.06 \\
   \hline
   \end{tabular}\\[1ex]
   \caption{Total solution times for the multigrid solvers in the DUNE and hypre libraries and for the geometric multigrid code. The AMG algorithms were used as preconditioners for CG, whereas the geometric multigrid algorithm was used as a stand-alone solver. All times are given in seconds.}
   \label{tab:MultigridTotalTimes}
  \end{center}
\end{table*}

For both AMG preconditioners the number of iterations is stable at around $9-13$. The time per iteration scales very well for the AMG preconditioners, in particular for the DUNE AMG solver. The parallel efficiency for BoomerAMG drops on 65536 cores. 
Further experiments with different problem sizes (not shown here) indicate that this is not an intrinsic problem of BoomerAMG (which has been shown to scale to larger core counts for different problems), but rather due to the fact that the ratio of horizontal to vertical grid spacing for the larger problem sizes becomes smaller in our scaling tests. As a consequence, the direction of the anisotropy changes within one vertical column, as can be seen by comparing $\beta_{\operatorname{bottom}}$ and $\beta_{\operatorname{top}}$ in Tab. \ref{tab:ParameterSpace}, leading to a vastly different coarsening strategy and a higher cost per iteration.

Note that this is not a problem for the geometric multigrid solver as the anisotropy is still grid-aligned. For the geometric multigrid solver the number of iterations is smaller than for either of the AMG methods and it is stable at 6 for all problem sizes, as can be seen from Tab. \ref{tab:MultigridTimePerIteration}. The time per iteration is also reduced by a factor of two relative to the AMG solvers. Taking into account the coarse grid setup times, the geometric multigrid solver is roughly a factor 10-20 faster than then the DUNE-ISTL and hypre solvers (see Tab. \ref{tab:MultigridTotalTimes}). The geometric multigrid solver is also more than 5 times faster than the one level method, since it requires about 7 times less iterations and each iteration is only 30-50\% more expensive.

\begin{figure}
 \begin{center}
  \includegraphics[width=\linewidth]{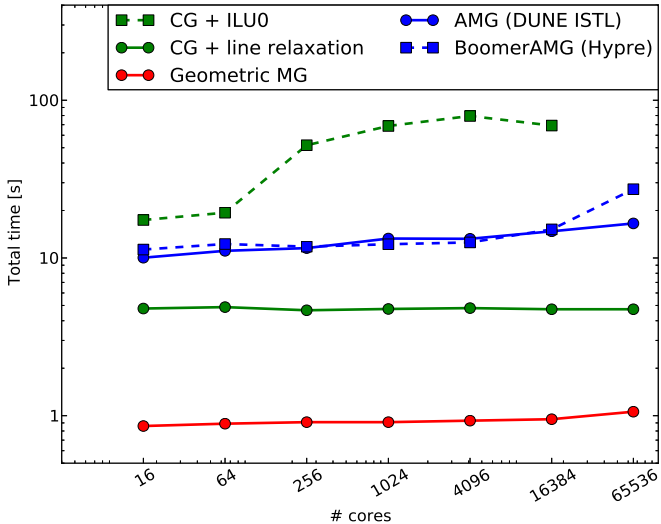}
  \caption{Weak scaling of the total solution times for the CG and multigrid solvers. In the case of AMG, the coarse grid setup time is included.}
  \label{fig:TotalTimeComparison}
 \end{center}
\end{figure}

The total solution times of all solvers are compared in Fig. \ref{fig:TotalTimeComparison}. All methods show good weak scaling. The geometric multigrid solver gives the best overall performance.
\subsubsection{Reduced number of multigrid levels}
\label{sec:ResultsGeoMG}
As remarked in sections \ref{sec:HorizontalCoupling} and \ref{sec:ImplementationGeometricMultigrid}, the conditioning of the problem improves with every coarsening step and reducing the total number of multigrid levels is expected to improve the parallel scalability. To confirm this we carried out weak scaling tests with reduced numbers of levels using the following setup:
\begin{itemize}
  \item \textit{Shallow multigrid.} The grid is coarsened until only one vertical column \textit{per core} is left. In our case, this corresponds to 7 multigrid levels. One application of the smoother is used on the coarsest level.
  \item \textit{Very shallow multigrid.} The grid is coarsened only three times. In contrast to the \textit{Standard} and \textit{Shallow} multigrid variant, the smoother is applied five times on the coarsest level.
\end{itemize}
While for the standard multigrid (i.e. coarsening until one global column is left) it is necessary to pull data together on the coarser processors, this is not necessary for the shallow or the very shallow multigrid setup. 

The results are shown in Tab. \ref{tab:GeometricMGPerformance}. They should be compared to those from the standard multigrid solver in Tab. \ref{tab:MultigridTimePerIteration} and for the preconditioned CG algorithm in Tab. \ref{tab:WeakScalingOneLevel}. Reducing the number of multigrid levels to four does not increase the number of iterations and improves the parallel scalability. 
\begin{table*}
\renewcommand{\arraystretch}{1.25}
\begin{center}
   \begin{tabular}{|rr|rrr|rrr|rrr|rrr|}
   \hline
   & & \multicolumn{3}{|c|}{shallow multigrid} & 
       \multicolumn{3}{|c|}{very shallow multigrid} \\
   \# cores  ($\mathcal{P}$) & \# dof & 
    \# iter & \titer & $E_S(\mathcal{P})$ & 
    \# iter & \titer & $E_S(\mathcal{P})$ \\
   \hline\hline
   16 & $8.3\cdot10^{6}$ & 
      6 & 0.143 & --- &
      6 & 0.143 & --- \\
   64 & $3.4\cdot10^{7}$ & 
      6 & 0.147 & 97\% &
      6 & 0.146 & 97\%  \\
   256 & $1.3\cdot10^{8}$ & 
      6 & 0.150 & 95\% &
      6 & 0.150 & 95\% \\
   1024 & $5.4\cdot10^{8}$ & 
      6 & 0.151 & 94\% &
      5 & 0.151 & 94\% \\
   4096 & $2.1\cdot10^{9}$ & 
      6 & 0.155 & 92\% &
      5 & 0.154 & 93\% \\
   16384 & $8.6\cdot10^{9}$ & 
      6 & 0.156 & 92\% &
      5 & 0.156 & 91\% \\
   65536 & $3.4\cdot10^{10}$ & 
      6 & 0.167 & 86\% &
      5 & 0.157 & 93\% \\
   \hline
   \end{tabular}\\[1ex]

  \caption{Number of iterations, time per iteration and scaled parallel efficiency for different numbers of multigrid levels $L$ in the geometric multigrid solver ($n_\text{loc} = 2^{19}$, $\mathcal{P}_0=16$).}
  \label{tab:GeometricMGPerformance}
\end{center}
\end{table*}
\subsubsection{Robustness.}\label{sec:Robustness}
Reducing the number of levels can, however, have an impact on the robustness of the method. To quantify this we investigate the dependency of the number of iterations for the geometric multigrid code under variations of the two model parameters $\omega^2$ and $\lambda^2$. We increase $\omega^2$ by a factor of $f_{\omega^2}=10$ and $100$ relative to the reference value, and vary $\lambda^2$ by a factor of $f_{\lambda^2}=10^2$ and $10^{-2}$. The results are shown in Tab. \ref{tab:RobustnessNumberOfIterations}, both for a relatively small problem and for a large problem (the same number of processors was used as for the runs in \ref{tab:GeometricMGPerformance}). 
\begin{table*}
\renewcommand{\arraystretch}{1.25}
  \begin{center}
   \begin{tabular}{|rr|rr|rr|rr|rr|}
    \hline
    \multicolumn{2}{|c|}{$3.4\cdot10^{7}$ dof} &
    \multicolumn{2}{|r|}{standard multigrid} & 
    \multicolumn{2}{|r|}{shallow multigrid} & 
    \multicolumn{2}{|r|}{very shallow multigrid} &
    \multicolumn{2}{|r|}{CG with line relaxation}\\
   $f_{\omega^2}$ & $f_{\lambda^2}$ & 
    \# iter & $t_{iter}$ & 
    \# iter & $t_{iter}$ & 
    \# iter & $t_{iter}$ & 
    \# iter & $t_{iter}$ \\
   \hline\hline
$1$ & $1$ & 
	 6 & 0.147 & 
	 6 & 0.148 & 
	 6 & 0.147 & 
	 43 & 0.113 \\
$1$ & $10^{2}$ & 
	 6 & 0.148 & 
	 6 & 0.147 & 
	 6 & 0.167 & 
	 42 & 0.113 \\
$1$ & $10^{-2}$ & 
	 6 & 0.148 & 
	 6 & 0.146 & 
	 6 & 0.147 & 
	 42 & 0.113\\
$10$ & $1$ & 
	 6 & 0.147 & 
	 6 & 0.148 & 
	 15 & 0.147 & 
	 140 & 0.112 \\
$100$ & $1$ & 
	 8 & 0.147 & 
	 10 & 0.147 & 
	 100 & 0.146 & 
	 300 & 0.112\\
\hline
   \end{tabular}\\[1ex]
   \begin{tabular}{|rr|rr|rr|rr|rr|}
    \hline
    \multicolumn{2}{|c|}{$8.6\cdot10^{9}$ dof} &
    \multicolumn{2}{|r|}{standard multigrid} & 
    \multicolumn{2}{|r|}{shallow multigrid} & 
    \multicolumn{2}{|r|}{very shallow multigrid} &
    \multicolumn{2}{|r|}{CG with line relaxation}\\
   $f_{\omega^2}$ & $f_{\lambda^2}$ & 
    \# iter & $t_{iter}$ & 
    \# iter & $t_{iter}$ & 
    \# iter & $t_{iter}$ & 
    \# iter & $t_{iter}$ \\
   \hline\hline
$1$ & $1$ & 
	 6 & 0.159 & 
	 6 & 0.155 & 
	 5 & 0.152 & 
	 40 & 0.117 \\
$1$ & $10^{2}$ & 
	 6 & 0.159 & 
	 6 & 0.153 & 
	 5 & 0.152 & 
	 39 & 0.117\\
$1$ & $10^{-2}$ & 
	 6 & 0.159 & 
	 6 & 0.154 & 
	 5 & 0.152 & 
	 37 & 0.118\\
$10$ & $1$ & 
	 6 & 0.159 & 
	 6 & 0.154 & 
	 14 & 0.173 & 
	 131 & 0.120 \\
$100$ & $1$ & 
	 6 & 0.161 & 
	 12 & 0.153 & 
	 129 & 0.226 & 
	 438 & 0.142\\
\hline
   \end{tabular}\\[1ex]
  \caption{Number of iterations and time per iteration for different parameter settings and solvers. For each problem size the reference values for $\omega^2$ and $\lambda^2$ in Tab. \ref{tab:ParameterSpace} were multiplied by the factors $f_{\omega_2}$ and $f_{\lambda^2}$. The runs with $3.4\cdot10^{7}$ degrees (top) of freedom were carried out on 64 processors and the runs with $8.6\cdot10^{9}$ degrees of freedom (bottom) on 16384 processors.}
  \label{tab:RobustnessNumberOfIterations}
  \end{center}
\end{table*}

For all solvers, the rate of convergence is independent of the vertical coupling parameter $\lambda^2$, as one would expect from the tensor product multigrid theory in \cite{BoermHiptmair1999}. 
However,  the multigrid variants with limited numbers of levels are affected by an increase in the time step size (and thus in $\omega^2$).  While the standard multigrid seems to be largely unaffected, the number of iterations increases as the total number of levels is reduced. This is particularly pronounced for the very shallow multigrid code and for the one-level method. 
\subsection{Strong scaling.}
\begin{figure}
  \includegraphics[width=1.0\linewidth]{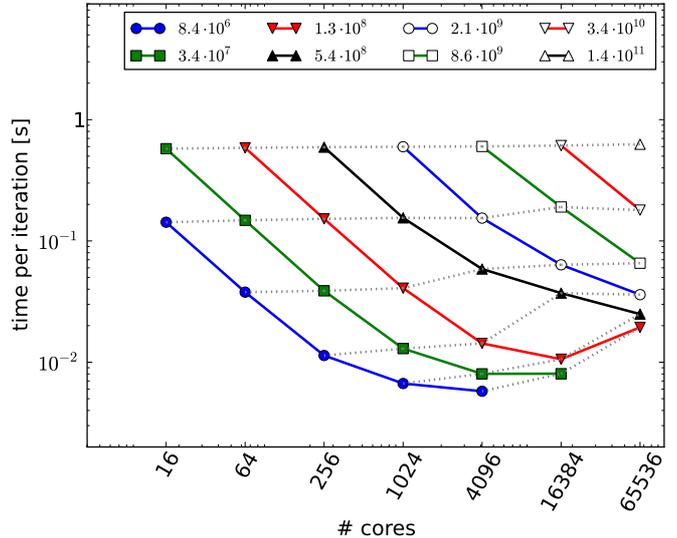}
  \caption{Strong scaling of the geometric multigrid code for different problem sizes. The time per iteration is shown as a function of the number of cores. The horizontal gray lines correspond to weak scaling experiments with different local problem sizes.}
  \label{fig:StrongScalingTimePerIteration}
\end{figure}

For the geometric multigrid code we also investigated strong scaling for different problem sizes, i.e. parallel speedup for a fixed problem size.
The time per iteration for different problem sizes is plotted in Fig. \ref{fig:StrongScalingTimePerIteration}. 

\begin{figure}
  \includegraphics[width=1.0\linewidth]{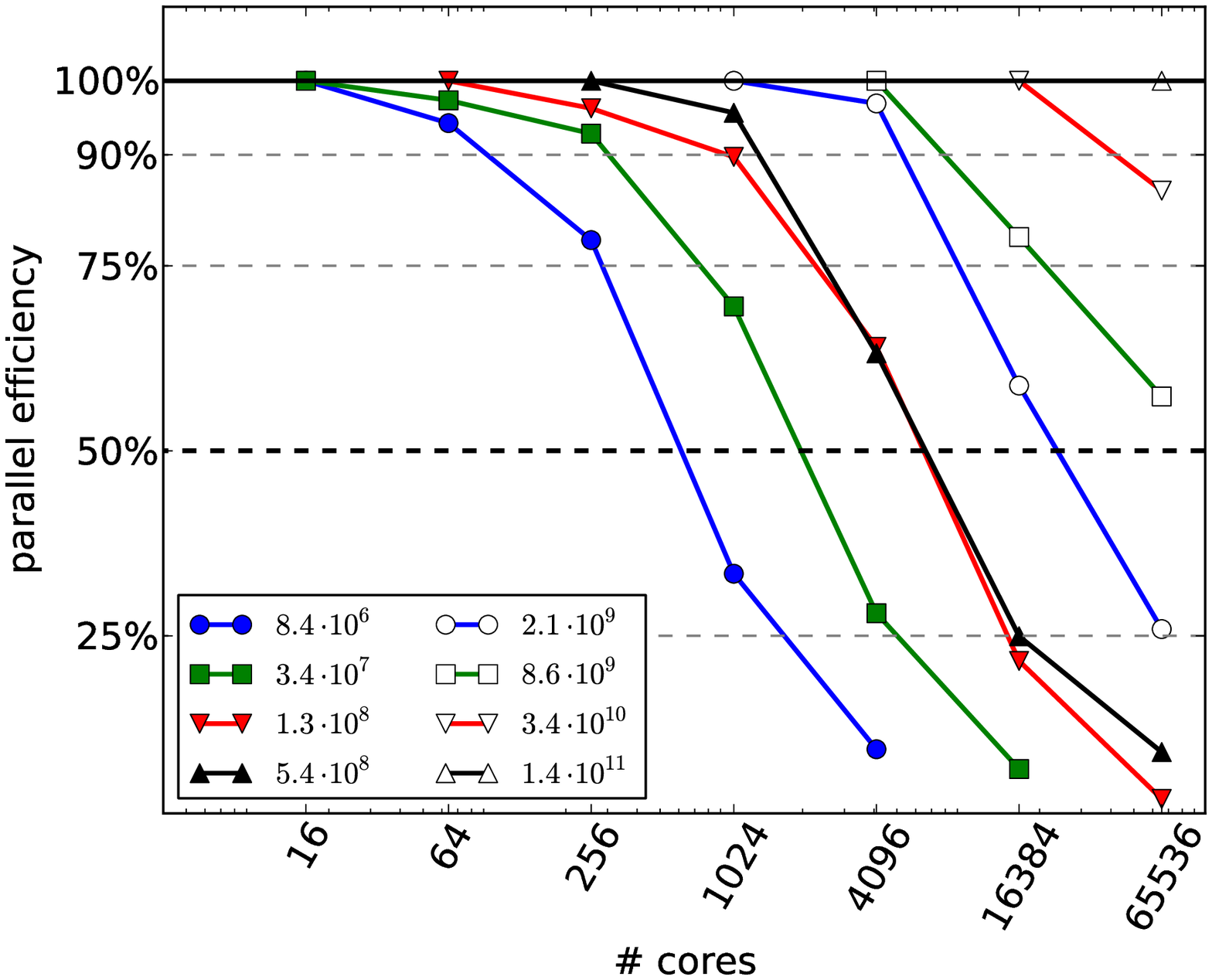}
  \caption{Strong scaling of the geometric multigrid code for different problem sizes. The parallel efficiency $E(\mathcal{P})$ is shown as a function of the number of cores $\mathcal{P}$.}
  \label{fig:StrongScalingParallelEfficiency}
\end{figure}

For each strong scaling experiment the parallel efficiency is defined as
\begin{eqnarray}
  E(\mathcal{P}) &=& \frac{\titer(\mathcal{P}_0; n) * \mathcal{P}_0}{\titer(\mathcal{P}; n) * \mathcal{P}}
\end{eqnarray}
where in each case $\mathcal{P}_0$ is the smallest number of processors used for solving a problem with $n$ degrees of freedom. This quantity is plotted in Fig. \ref{fig:StrongScalingParallelEfficiency}.
For all problem sizes parallel
efficiency drops below $50\%$ for problems which have $8 \times 8 = 64$ or less vertical columns per processor. The latency of HECToR internode communications is around $1\mu s$ and the theoretical peak performance of the 90,112 machine is quoted as 800 TFLOPs. This implies that communication costs can only be ``hidden'' by overlapping them with calculations if at least around 10000 floating point operations are carried out per halo exchange. In particular strong scaling will start to break down once the number of operations per halo exchange drops below this limit. Assuming that around 10-100 floating point operations are required per grid cell, we expect this to be the case on the finest multigrid level once the problem size is reduced to 100-1000 cells (1-10 vertical columns). On the coarser multigrid levels, which account for a smaller fraction of the total runtime, this will occur earlier, so for the multigrid algorithm we expect strong scaling to break down for slightly larger problem sizes, as observed in our numerical results. In addition it should be kept in mind that in practice the latency will be much larger, especially for runs with a large number of cores, so that this number should only be regarded as a theoretical lower limit.
\subsection{Implementation on the entire sphere}
\label{sec:ResultsDUNEGrid}
All runs reported in the previous sections were carried out on one logically rectangular panel of the cubed sphere grid. We also implemented a geometric multigrid solver on a grid covering the entire sphere in DUNE. For this, a two dimensional host grid is created for the surface of the sphere and on each two-dimensional grid element a vector of length $n_z$ representing the vertical degrees of freedom, is stored. While data is addressed indirectly in the horizontal direction, it is stored consecutively in memory and addressed directly in the vertical direction.

While the implementation in DUNE is straightforward, currently the performance of the code is limited by the underlying grid implementation. The only grid which was found to give reasonable results is \texttt{UGGrid} \cite{Bastian1997}. However, currently the scalability of this implementation is limited and we are working with the DUNE developers to extend it to larger core counts.

Weak scaling results obtained on a full cubed sphere grid are shown in Tab. \ref{tab:ResultsDUNEGrid}. The number of multigrid levels is 6 in each case and one iteration of block-Jacobi line relaxation is used for pre- and post-smoothing. Note that the number of degrees of freedom per core is a factor of four smaller than in the runs reported in the previous sections. The number of iterations is larger as Jacobi relaxation is a less efficient smoother than RB SOR, but as for the runs on one panel of the cubed sphere the iteration count does not increase with the model resolution.
More detailed results will be reported in a forthcoming publication (\cite{Dedner2012}).
\begin{table}
\renewcommand{\arraystretch}{1.25}
\begin{center}
   \begin{tabular}{|rr|rrr|}
   \hline
   \# cores ($\mathcal{P}$)& \# dof & 
    \# iter & \titer & $E_S(\mathcal{P})$ \\
   \hline\hline
   6  & $7.9\cdot10^{5}$ & 
      12 & 0.313 & --- \\
   24  & $3.1\cdot10^{6}$ & 
      12 & 0.387 & $81\%$ \\
   96  & $1.3\cdot10^{7}$ & 
      11 & 0.374 & $84\%$ \\
   384  & $5.0\cdot10^{7}$ & 
      11 & 0.411 & $76\%$ \\
   \hline
   \end{tabular}\\[1ex]
  \caption{Number of iterations, time per iteration and scaled parallel efficiency for the geometric multigrid solver on a full cubed sphere grid, implemented in DUNE, based on \texttt{UGGrid} with $n_\text{loc} = 2^{17}$ and $\mathcal{P}_0 = 6$. Block-Jacobi smoothing was used with 1 pre- and 1 post-smoothing step and the number of multigrid levels was $6$ in all runs.}
  \label{tab:ResultsDUNEGrid}
\end{center}
\end{table}
\section{Conclusions}\label{sec:Conclusions}
In this article we discussed efficient and scalable solvers for the elliptic PDE arising from semi-implicit semi-Lagrangian time stepping in the dynamical core of numerical weather- and climate- prediction models.

After reviewing modern iterative solvers, in particular suitably preconditioned Krylov subspace and multigrid methods, as well as the  existing literature on their application and parallel scaling in NWP applications, we reported on the results of massively parallel scaling tests for a model equation. An important characteristic of this equation is the strong coupling in the vertical direction and the presence of a zero order term, which leads to a well conditioned problem on coarser multigrid levels.

We tested and optimised algebraic multigrid solvers in existing libraries (DUNE and hypre) and developed a bespoke geometric multigrid algorithm based on the tensor product idea in \cite{BoermHiptmair1999} that is well-suited to the strong vertical anisotropy. The bespoke solver avoids matrix- and coarse level setup costs by recalculating the matrix stencil on the fly. We compared the multigrid solvers to various implementations of preconditioned CG.  All solvers show good weak scaling on up to 65536 cores of the HECToR supercomputer. In comparison to the one-level Krylov subspace methods, the number of iterations and the overall computational time can be reduced significantly with the use of multigrid methods. Here, we find that the geometric multigrid method is superior to the AMG solvers both in terms of the time per iteration and the number of iterations and it is possible to solve the model equation with $3.4\cdot 10^{10}$ degrees of freedom in around one second on 65536 processors.  In contrast to one-level methods, the multigrid solvers are robust with respect to variations of the model parameters. Finally, as the problem is well conditioned on coarser levels, it is not necessary to coarsen the problem down to one global column. Using a reduced number of coarse levels can improve the parallel scalability, but it also affects the robustness with respect to the time step size. For time step sizes typically encountered in operational runs this does not appear to be a problem though.

We conclude that in contrast to common misconceptions, the elliptic solve in each time step does not limit the scalability of implicit or semi-implicit methods in NWP. Relative to explicit (or vertically-implicit) methods, semi-implicit time stepping is considered to be more robust and it allows for a larger model time step, which is why several of the current operational forecasting centres (such as the UK Met Office or the ECMWF) use it. The current paper was intended to show that these advantages do not have to be forfeit for better parallel scalability on future, massively parallel architectures. It remains to be seen, however, which of the two methods (semi-implicit or explicit) leads to the shortest total runtime in a real simulation. This is beyond the scope of this article and requires further investigations for a particular model implementation. 

There are several ways of further improving on the current work: while the strong scaling results reported here look promising, there is room for improvement, for example by using a hybrid MPI/OpenMP implementation. Furthermore, the runs reported here were carried out on 1/6th of a cubed sphere grid, but we also implemented a geometric multigrid code for arbitrary horizontal grids on the sphere. This is the subject of a forthcoming publication \cite{Dedner2012}.
\section{Acknowledgements}
This work was funded as part of the NERC project on ``Next Generation Weather and Climate Prediction'' (NGWCP), grant number NE/J005576/1. We would like to thank the dynamics research group at the Met Office and all members of the GungHo! project for useful discussions. We thank the DUNE developers and in particular Dr. Andreas Dedner and the group of Prof. Peter Bastian in Heidelberg for help with DUNE related questions and for their hospitality during EM's stay in Heidelberg in December 2012. Similarly, we are grateful for the support of the hypre developers, in particular Dr. Robert Falgout and Dr. Ulrike Maier-Yang, with optimising the BoomerAMG solver, as well as their hospitality during RS and EM's stay at LLNL in July 2012. This work made use of the facilities of HECToR, the UK's national high-performance computing service, which is provided by UoE HPCx Ltd at the University of Edinburgh, Cray Inc and NAG Ltd, and funded by the Office of Science and Technology through EPSRC's High End Computing Programme.
\ifpreprint
\appendix
\section{Detailed derivation of elliptic equation for pressure correction}
\label{sec:DerivationModelEquation}
In this section we show how the elliptic PDE (\ref{eqn:FullSISLEquation}) for the pressure correction in semi-implicit semi-Lagrangian time stepping can be derived from the fundamental equations of atmospheric motion and motivate the model equation (\ref{eqn:ModelEquation}) and the size of the parameters $\omega^2$ and $\lambda^2$. The derivation is the continuum equivalent of the derivation in \cite{Wood2013} (see also \cite{Melvin2010}), where the construction of the ENDGame dynamical core of the Met Office Unified core is outlined.
\subsection{Fundamental equations}
Ignoring the Coriolis force, the fundamental equations of atmospheric flow for dry air are given by
\begin{eqnarray}
  \DDt{\vec{u}} &=& -c_p \theta \nabla \pi + \vec{S}^{\vec{u}} \quad\text{(momentum equation)},\label{eqn:fund_momentum}\\[1ex]
  \DDt{\theta} &=& S^{\theta} \qquad\text{(thermodynamic equation)}\label{eqn:fund_thermodynamic},\\[1ex]
  \DDt{\rho} &=& -\rho \nabla \cdot \vec{u} \qquad\text{(continuity equation)},\label{eqn:fund_continuity}\\[1ex]
  \theta \rho &=& \Gamma \pi^{\gamma} \qquad\text{(ideal gas law)}
  \label{eqn:fund_idealgas}
\end{eqnarray}
with 
\begin{xalignat}{3}
  \Gamma &\equiv p_0/R_d, &
  \kappa &\equiv R_d/c_p, &
  \gamma &\equiv \frac{1-\kappa}{\kappa}. \label{eqn:Definitiongamma}
\end{xalignat}
Here $p_0$ is a reference pressure; $c_p$ and $R_d$ are the specific heat capacity and specific gas constant of dry air.
We use the numerical values of
\begin{xalignat}{2}
  R_d &= 286.9 \operatorname{J}\operatorname{kg}^{-1}\operatorname{K}^{-1}, &
  c_p &= 1006.0 \operatorname{J}\operatorname{kg}^{-1}\operatorname{K}^{-1},\notag\\
  \kappa &= 0.2852, &
  \gamma &= 2.506.\label{eqn:AtmosphericConstants}
\end{xalignat}

The external forcings $S^X$ represent the model physics below the grid scale and the gravitational acceleration has been absorbed into $\vec{S}^{\vec{u}}$.
Potential temperature $\theta$ and Exner pressure $\pi$ are related to canonical temperature and pressure via
\begin{xalignat}{2}
  \pi &= \left(p/p_0\right)^{\kappa}, &
  \theta &= T \left(p/p_0\right)^{-\kappa} = T/\pi.
\end{xalignat}
\subsection{Pressure correction equation with background profiles}
In the following the fundamental equations in (\ref{eqn:fund_idealgas}) are linearised around background profiles, denoted by superscript ${}^\bg$.

Write $\theta = \theta^\bg+\theta'$, $\rho=\rho^\bg+\rho'$ and $\pi=\pi^\bg+\pi'$ (as well as $\vec{u}=\vec{u}'$ and $w=w'$, but the prime on the velocities are not written out in the following). Assume that the profiles $\theta^\bg$, $\rho^\bg$ and $\pi^\bg$ depend on the vertical coordinate only and are constant in time. In this section we always write $\vec{u}$ for the two dimensional horizontal velocity, $w$ for the vertical component of the velocity and $\nablahoriz$ for the horizontal derivative tangential to the surface of the earth; $\laplhoriz$ denotes the horizontal Laplacian. We also linearise the equations by dropping terms which contain products of two or more primed quantities (or assume implicitly that they are moved to the right hand side, if a non-incremental iteration scheme is used to solve the non-linear equation).

The fundamental equations (\ref{eqn:fund_momentum}), (\ref{eqn:fund_thermodynamic}) and (\ref{eqn:fund_continuity}) become
\begin{eqnarray}
  \DDt{\vec{u}} &=& c_p \theta\frac{\nablahoriz \pi'}{r} + \vec{S}_{\vec{u}}, \\[1ex]
  \DDt{w} &=& -c_p \theta \ddr{\pi} + S_w, \\[1ex]
  \DDt{\theta'} + w\ddr{\theta^\bg} &=& S_\theta, \\[1ex]
  \DDt{\rho'} + w\ddr{\rho^\bg} &=& -\rho \left(\frac{\nablahoriz\cdot\vec{u}}{r}+\frac{1}{r^2}\frac{\partial}{\partial r}\left(r^2 w\right)\right),\notag\\[1ex]
\end{eqnarray}
and the linearised ideal gas law (\ref{eqn:fund_idealgas}) is
\begin{eqnarray}
  \frac{\rho'}{\rho^\bg}+\frac{\theta'}{\theta^\bg} &=& \gamma\frac{\pi'}{\pi^\bg}.
\end{eqnarray}
\subsection{Semi-implicit semi-Lagrangian timestepping}
Using the semi-implicit semi-Lagrangian time discretisation in (\ref{eqn:SISLScheme}) we obtain a coupled set of equations for the fields at the next time step (in the following $\vec{u} = \vec{u}^{(n+1)}$ etc. to simplify the notation):
\begin{eqnarray}
  \vec{u} &=& \vec{R}_{\vec{u}} - \adt\;c_p \theta^\bg \frac{\nablahoriz\pi'}{r}
  \label{eqn:derivationBG_SISL_horizmom} \\[1ex]
  w &=& R_w - \adt\; c_p \theta \ddr{\pi}\\[1ex]
  \theta' &=& R_\theta - \adt\; w\ddr{\theta^\bg}\\[1ex]
  \rho' &=& R_\rho - \adt\Big[w\ddr{\rho^\bg} 
\label{eqn:derivationBG_SISL_continuity} 
\\&&\qquad+\;\; \rho^\bg\left(\frac{\nablahoriz\cdot\vec{u}}{r}+\frac{1}{r^2}\frac{\partial}{\partial r}\left(r^2 w\right)\right)\Big]\notag
\end{eqnarray}
where we replaced $\theta$ by $\theta^\bg$ and $\pi$ by $\pi^\bg$ wherever possible.
Eliminating the horizontal velocity $\vec{u}$ is straightforward: take the divergence of the first equation and insert it in the last equation to obtain, after division by $\rho^\bg$
\begin{eqnarray}
  \frac{\rho'}{\rho^\bg} &=& \frac{R_\rho- \adt \;\rho^\bg\tfrac{1}{r} \nablahoriz\cdot\vec{R}_\vec{u}}{\rho^\bg}  - \adt\frac{1}{\rho^\bg}\ddr{\rho^\bg}w\notag\\[1ex]
    &&\quad+\;\;(\adt)^2 c_p\theta^\bg\frac{\laplhoriz\pi'}{r^2}-\adt\;\frac{1}{r^2}\frac{\partial}{\partial r}\left(r^2 w\right)\notag\\[1ex]
    \label{eqn:derivationBG_densitycorrection}
\end{eqnarray}
Again $\rho$ has been replaced by $\rho^\bg$ where possible.
The vertical velocity $w$ and $\theta'$ can be expressed in terms of the pressure correction by using the two remaining equations after linearisation,
\begin{eqnarray}
  \theta' &=& R_\theta -\adt\; \ddr{\theta^\bg} w \\[1ex]
  w &=& R_w - \adt\; c_p\left(\theta^\bg\ddr{\pi'}+\ddr{\pi^\bg}\theta'\right)
\end{eqnarray}
which give
\begin{eqnarray}
  w &=& \Lambda^\bg \left(f_2-\adt\; c_p \theta^\bg\ddr{\pi'}\right),\label{eqn:wthetarpelacement}\\[1ex]
  \theta' &=& \Lambda^\bg\left(f_3+(\adt)^2c_p\theta^\bg\ddr{\theta^\bg}\ddr{\pi'}\right)\notag.
\end{eqnarray}
The buoyancy frequency $\left(N^\bg\right)^2$ in
\begin{eqnarray}
\Lambda^\bg &=& \left[1+(\adt)^2\left(N^\bg\right)^2\right]^{-1}
\end{eqnarray}
is given by\footnote{$g$ is negative and $\theta^\bg$ decreases with height, so that $\left(N^\bg\right)^2$ is positive.}
\begin{eqnarray}
  \left(N^\bg\right)^2 &=& -c_p\ddr{\pi^\bg}\ddr{\theta^\bg} = \frac{g}{\theta^\bg}\ddr{\theta^\bg}\label{eqn:definitionbuoyancyfrequency}
\end{eqnarray}
and varies with height. The functions $f_2$ and $f_3$ are defined as
\begin{eqnarray}
  f_2 &=& R_w - \adt\; c_p \ddr{\pi^\bg} R_\theta,\\[1ex]
  f_3 &=& R_\theta - \adt \ddr{\theta^\bg}R_w.\notag
\end{eqnarray}
Next, use the linearised ideal gas law to replace $\rho'/\rho^\bg$ in (\ref{eqn:derivationBG_densitycorrection}) by $\theta'/\theta^\bg$ and $\pi'/\pi^\bg$. Finally, eliminate $\theta'$ and $w$ with the help of (\ref{eqn:wthetarpelacement}) and multiply by $\pi^\bg$ to obtain a differential equation for the pressure correction:
\begin{eqnarray}
  -(\adt)^2c_p\theta^\bg\pi^\bg\left(\laplhoriz
  + D^{(z)}
   \right)\pi'+\gamma \pi' &=& f.\notag\\
   \label{eqn:eqn:derivationBG_HelmholtzOperator}
\end{eqnarray}
The derivative operator in the vertical direction is defined as
\begin{eqnarray}
D^{(z)}X = {\frac{1}{\left(\pi^\bg\right)^\gamma c_p\theta^\bg r^2}}\,\frac{\partial}{\partial r}\left(\frac{\left(\pi^\bg\right)^\gamma c_p\theta^\bg {r^2}}{1+(\adt)^2\left(N^\bg\right)^2}\frac{\partial X}{\partial r}\right)\notag.\\[2ex]
\label{eqn:verticalderivative}
\end{eqnarray}
\subsection{Reference profiles}
Suitable background profiles have to be chosen for $\theta^\bg$, $\pi^\bg$ and $\rho^\bg$ in (\ref{eqn:eqn:derivationBG_HelmholtzOperator}), (\ref{eqn:verticalderivative}), one could for example choose an isothermal profile. In the following we assume that the profiles are constant, for simplicity. For isothermal profiles and neglecting the vertical variation of gravity, the buoyancy frequency is constant and given by
\begin{eqnarray}
  N^\bg &=& g_0^2/c_h^2 = 0.018\Us^{-1}.
\end{eqnarray}
For the numerical experiments in this article we consider the following simple model problem, which captures the relative size of the second order term ($\omega^2$) and the strength of the vertical coupling ($\lambda^2$):
\begin{eqnarray}
  -\omega^2 \left(\laplhoriz u+\lambda^2  \frac{1}{r^2}\frac{\partial}{\partial r}\left( r^2 \frac{\partial u}{\partial r}\right)\right)+u &=& f\notag\\[0ex]
  \label{eqn:ModelEquationAppendix}
\end{eqnarray}
To be consistent with the mathematical literature $u$ denotes the Exner pressure increment $\pi'$ and $\laplhoriz$ is the two dimensional Laplace operator. By comparison to the physical problem, and rescaling to dimensionless coordinates of $\order{1}$ the coefficients $\omega^2$ and $\lambda^2$ are given by
\begin{xalignat}{2}
  \omega^2 &\approx \left(\frac{c_h\adt}{\Rearth}\right)^2 &
  \lambda^2 &\approx \frac{1}{1+(\adt)^2\left(N^\bg\right)^2}
\end{xalignat}
With the numerical values in (\ref{eqn:AtmosphericConstants}) we obtain a typical speed of $c_h=\sqrt{\gamma} c_s=550ms^{-1}$ where $c_s=\sqrt{c_pT_0/\gamma}$ is the speed of sound at ground level.

The problem is solved in the spherical domain defined by $1\le x^2+y^2+z^2\le1+H$ where $H=D/\Rearth$. Here $\Rearth$ is the radius of the earth and $D$ the thickness of the atmosphere.
\section{Algorithms}
\label{sec:Algorithms}
In the following we give explicit forms of some of the algorithms discussed in the main text.

Algorithm \ref{alg:PCG} is the Preconditioned Conjugate Gradient (PCG) method for solving $Au=f$ with the preconditioner $M^{-1}$. The iteration stops as soon as the maximal number of iterations ($\operatorname{maxiter}$) has been reached, the residual has been reduced by a factor $\epsilon$ or it is smaller than some tolerance $\tau$.
\begin{algorithm}
  \caption{Preconditioned Conjugate Gradient method}
  \label{alg:PCG}
  \begin{algorithmic}[1]
    \STATE $r_0\mapsfrom f-Au_0$ \COMMENT{Calculate residual}
    \STATE $z_0 = M^{-1}r_0$ \COMMENT{Apply preconditioner}
    \STATE $p_0\mapsfrom z_0$
    \STATE $\kappa_{old} \mapsfrom \langle r_0,z_0\rangle$
    \FOR{$K=1,\operatorname{maxiter}$}
      \STATE $q_{K-1} \mapsfrom Ap_{K-1}$ \COMMENT{Apply operator}
      \STATE $\alpha \mapsfrom \kappa_{old}/\langle p_{K-1},q_{K-1}\rangle$
      \STATE $u_{K}\mapsfrom u_{K-1}+\alpha p_{K-1}$ \COMMENT{Update solution}
      \STATE $r_{K}\mapsfrom r_{K-1}+\alpha q_{K-1}$ \COMMENT{Update residual}
      \IF{$||r_K||/||r_0|| < \epsilon$ or $||r_K||<\tau$}
        \STATE Exit \COMMENT{Stopping criterion}
      \ENDIF
      \STATE $z_K = M^{-1}r_K$ \COMMENT{Apply preconditioner}
      \STATE $\kappa\mapsfrom \langle r_K,z_K\rangle$
      \STATE $\beta\mapsfrom \kappa/\kappa_{old}$
      \STATE $p_K\mapsfrom z_{K}+\beta p_{K-1}$ \COMMENT{Update search direction}
      \STATE $\kappa_{old}\mapsfrom \kappa$
    \ENDFOR
    \STATE $u\mapsfrom u^{(K)}$
    \RETURN $u$
  \end{algorithmic}
\end{algorithm}
The tridiagonal matrix (or Thomas-) algorithm for the solution of the $n_z$ dimensional tridiagonal system 
\begin{eqnarray}
  \begin{pmatrix}
    a_1 & b_1 &  \\
    c_2 & \ddots & \ddots  \\
      & \ddots & \ddots
      & b_{n_z-1} \\
     & & c_{n_z} & a_{n_z}
  \end{pmatrix}
  \begin{pmatrix}
    u_1 \\
    u_2 \\
    \ddots\\
    u_{n_z-1}\\
    u_{n_z}
  \end{pmatrix}
&=&
  \begin{pmatrix}
    f_1 \\
    f_2 \\
    \ddots\\
    f_{n_z-1}\\
    f_{n_z}
  \end{pmatrix}\notag
  \\[2ex]
\end{eqnarray}
is shown in Algorithm \ref{alg:ThomasAlgorithm}. The algorithm consists of two steps: after construction of the modified coefficient vectors $b'$ and $f'$, backward substitution is used to obtain the solution vector $u$. The computational cost of the algorithm is $\order{n_z}$.
\begin{algorithm}
  \caption{Tridiagonal matrix algorithm}
  \label{alg:ThomasAlgorithm}
  \begin{algorithmic}[1]
    \STATE $b'_1\mapsfrom b_1/a_1$
    \STATE $f'_1\mapsfrom f_1/a_1$
    \FOR{$i=2,\dots,n_z-1$}
      \STATE $b'_i\mapsfrom b_i/(a_i-b'_{i-1}c_{i})$
      \STATE $f'_i\mapsfrom (f_i-f'_{i-1}c_{i})/(a_i-b'_{i-1}c_{i})$
    \ENDFOR
    \STATE $f'_{n_z}\mapsfrom (f_{n_z}-f'_{n_z-1}c_{n_z})/(a_{n_z}-b'_{n_z-1}c_{n_z})$
    \STATE $u_{n_z}\mapsfrom f'_{n_z}$
    \FOR{$i=n_z-1,n_z-2,\dots,1$}
      \STATE $u_{i}\mapsfrom f'_i - b'_i u_{i+1}$
    \ENDFOR
    \RETURN $u$
  \end{algorithmic}
\end{algorithm}
\fi 
\ifpreprint 
\bibliographystyle{unsrt}
\else 
\bibliographystyle{wileyqj}
\fi 

\end{document}